\documentclass[preprint,12pt]{elsarticle}

\usepackage{amssymb}
\usepackage[cmex10]{amsmath}
\usepackage{amsfonts}

\usepackage{color}

\usepackage{graphicx}
\usepackage{graphicx,subfigure}
\usepackage{epstopdf}
\usepackage{soul}

\usepackage{calc}
\usepackage{amsmath}
\usepackage{xcolor}

\usepackage{lineno}

\journal{.}

\begin{document}

\begin{frontmatter}



\title{A global optimization paradigm based on change of measures\tnoteref{label1}}
\tnotetext[label1]{We thank Prof. K P J Reddy, Department of Aerospace Engineering, Indian Institute of Science, for sharing with us the experimental data}

\author{Saikat Sarkar\(^1\), Debasish Roy\(^1\)* and Ram Mohan Vasu\(^2\)}

\address{\(^1\)Computational Mechanics Lab, Department of Civil Engineering,
\par \(^2\)Department of Instrumentation and Applied Physics
Indian Institute of Science, Bangalore 560012, India
\par *Corresponding author; email: royd@civil.iisc.ernet.in
}

\begin{abstract}
A global optimization framework, acronymed COMBEO ({\bf{C}}hange {\bf{O}}f {\bf{M}}easure
{\bf{B}}ased {\bf{E}}volutionary {\bf{O}}ptimization), is proposed. An important aspect in the development
is a set of derivative-free additive directional terms obtainable through a change of measures \emph{en route} to the imposition of any
 stipulated conditions aimed at driving the realized design variables (particles) to the global
optimum. The generalized setting offered by the new approach also
 enables several basic ideas, used with other global search methods such as the
particle swarm or the differential evolution, to be rationally incorporated in the proposed setup via a change of
measures. The global search may be further aided by imparting to the directional update terms
 additional layers of random perturbations such as `scrambling' and `selection'. Depending on
 the precise choice of the optimality conditions and the extent of random perturbation, the
search can be readily rendered either greedy or more exploratory. As numerically demonstrated, the new
proposal appears to provide for a more rational, more accurate and faster alternative to
most available evolutionary optimization schemes, prominent amongst which are the differential evolution
 and the particle swarm methods.
\end{abstract}


\end{frontmatter}


\section{Introduction}
 Potential applications of global optimization are ubiquitous across a large spectrum of problems in science and engineering, from VLSI circuit design to optimization of transportation routes to determination of protein structures. An essential aspect of the solution to any such problem is the extremization of an objective functional, possibly subject to a prescribed set of constraints. The solution is deemed to have been achieved if the global extremum of the objective functional is reached. For many practical applications, the objective functional could be non-convex, non-separable and even non-smooth. This precludes gradient-based local optimization approaches to treat these problems, an aspect that has driven extensive research in devising global optimization strategies, many of which employ stochastic (i.e. random evolutionary) search of heuristic or meta-heuristic origin \cite{1holland}. Indeed, finding the global optimum within a search space of the parameters or design variables is much more challenging than arriving at a local extremum. In the absence of a generic directional information (equivalent to the Gateaux or directional derivative in the search for a local extremum of a sufficiently smooth cost functional), a direct search seems to be the only option left. Some of the notable schemes in this genre include variants of the genetic algorithm (GA) \cite{2goldberg_GA}, differential evolution (DE) \cite{3storn_DE}, particle swarm optimization (PSO) \cite{4kennedy_PSO} and ant colony optimization \cite{5dorigo_antcolony}. In a search for the global extremum, stochastic schemes typically score over their gradient-based deterministic counterparts \cite{6fletcher_opt, 7chong_zak}, even for cases involving sufficiently smooth cost functionals with well defined directional derivatives. Nevertheless, thanks to the proper directional information guiding the update, gradient-based methods possess the benefit of faster convergence to the nearest extremum, as long as the objective functional is sufficiently smooth. To the authors' knowledge, the existing evolutionary global optimization schemes do not offer the benefit of a well founded directional information. Most evolutionary schemes depend on a random (diffusion type) scatter applied to the available candidate solutions or particles and some criteria for selecting the new particles. Unfortunately, a `greedy' approach for the global optimum (e.g. selecting particles of higher `fitness' with higher probability), though tempting from the perspective of faster convergence, faces the hazard of getting trapped in local extrema. As a fix to a premature and possibly wrong convergence, many evolutionary global search schemes adopt safeguards. Despite the wide acceptability of a few evolutionary methods of the heuristic/meta-heuristic origin \cite{8glover_metaheuristic}, the underlying justification is often based on sociological or biological metaphors\cite{1holland, 2goldberg_GA, 3storn_DE, 4kennedy_PSO, 5dorigo_antcolony} that are hardly founded on a sound probabilistic basis even though a random search forms a key ingredient of the algorithm. Be that as it may, the popular adoption of these schemes is not only due to the algorithmic simplicity, but mainly because of their effectiveness in treating many Np-hard optimization problems, which may be contrasted with a relatively poorer performance of some of the well-grounded stochastic schemes, e.g. simulated annealing \cite{9SA_science}, stochastic tunneling \cite{10stochastic_tunneling} etc. Whether a slow convergence to the global optimum, not infrequently encountered with some of the popular evolutionary schemes, could be fixed by reorienting them with an alternative stochastic framework, however, remains a moot point, which is addressed in this work to an extent. Another related question is regarding the precise number of parameters in the algorithm that the end-user must tune for an appropriate `{\emph{exploration-exploitation trade-off}}', a phrase used to indicate the relative balance of random scatter of the particles vis-a-vis their selection based on the `fitness'. Admittedly, the notion of random evolution, built upon Monte Carlo (MC) sampling of the particles, efficiently explores the search space, though at the cost of possibly slower convergence \cite{11hybrid_opt} in contrast to a gradient based method. For better exploration, many such schemes, e.g. the GA, adopt steps like  `crossover', `mutation', `selection' etc. While `crossover' and `mutation' impart variations in the particles, by the `selection' step each particle is assigned a `weight' or fitness value (a measure of importance) determined as the ratio of the realized value of the objective functional to the available maximum of the same. In the selection step, the fitness values, used to update the particles via selection of the best-fit individuals for subsequent crossover and mutation, may be considered functionally analogous to the derivatives used in gradient based approaches. Analogy may also be drawn between the fitness values and the likelihood ratios commonly employed in non-linear stochastic filtering, e.g. the family of sequential importance sampling filters \cite{12Gordon_filt}. Even though a weight based approach reduces a measure of misfit between the available best and the rest within a finite ensemble of particles, they bring with them the curse of degeneracy, wherein all particles but one tend to receive zero weights as the iterations progress \cite{13arunampalam}. This problem, also referred to as `particle collapse' in the stochastic filtering parlance, can be arrested, in principle, by exponentially increasing the ensemble size (number of particles), a requirement that can hardly be met in practice \cite{14snyder_dimensionality}. Evolutionary schemes that replace the weight-based multiplicative approach by adopting additive updates for the particles, often succeed in eliminating particle degeneracy. Evolutionary schemes like DE, PSO etc. that are known to perform better than GA utilize such additive particle update strategies without adversely affecting the `craziness' or randomness in the evolution. On reaching the optimum, the additive correction shrinks to zero. Unfortunately, none of these methods obtain the additive term, which may be looked upon as a stochastic equivalent to the derivative-based directional information, in a rigorous or optimal manner - a fact that is perhaps responsible for a painstakingly large number of functional evaluations in some cases.

 \par  In framing a rigorous stochastic basis leading to directional updates of the additive type, several possible schemes are suggested in this work. First it is shown that the problem of optimization may be generically posed as a {\it{martingale problem}} \cite{15oksendal, 16stroock_varadhan}, which must however be randomly perturbed to facilitate the global search. For instance, one may perform a greedy search for an extremum of an objective functional by solving a martingale problem, which is in the form of an integro-differential equation herein referred to as the extremal equation. In line with the Freidlin-Wentzell theory \cite{17freidlin_wentzell}, the martingale problem may further be perturbed randomly so as to enable the greedy scheme to converge to the global extremum as the strength of the perturbation vanishes asymptotically. The solution through this approach, which is viewed as a stochastic process parametered via the iterations, depends crucially on an error or `innovation' function and is deemed to have been realized when the innovation function, interpreted as a stochastic process, is driven to a zero-mean noise or martingale \cite{15oksendal}. The martingale structure of the innovation essentially implies that the mean of the computed cost functional in future iterations remains constant and invariant to zero-mean random perturbations of the argument vector (the design variables). The argument vector should then correspond to an extremum. In the evolutionary random setup, both the cost functional and its argument vector are treated as stochastic diffusion processes even though the original extremization problem is deterministically posed. Realizing a zero-mean martingale structure for the innovation requires a change of measures that in turn leads to additive gain-type updates on the particles. The gain coefficient matrix, which is a replacement for and generalization over the Frechet derivative of a smooth cost functional, provides an efficacious directional search without requiring the functional to be differentiable. Additionally, in order to facilitate the global search, a general random perturbation strategy (using steps such as `scrambling', `relaxation' and `selection') is adopted to insure against a possible trapping of particles in local extrema.

\par An important feature of the present approach is the flexibility with which the innovation process may be designed in order to meet a set of conflicting demands {\it{en route}} the detection of the global extremum. The efficiency of the global search basically relies upon the ability of the algorithm to explore the search space whilst preserving some directionality that helps to quickly resolve the nearest extremum. The development of the proposed setup, referred to as {\bf{COMBEO}} ({\bf{C}}hange {\bf{O}}f {\bf{M}}easure {\bf{B}}ased {\bf{E}}volutionary {\bf{O}}ptimization), recognizes the near impossibility of a specific optimization scheme performing uniformly well across a large class of problems. Accommodation of this fact had earlier led to a proposal of an evolutionary scheme that simultaneously ran different optimization methods for a given problem with some communications built amongst the updates by the different methods \cite{24multimethod_search}. A major related contribution of this work is to bring the basic ideas for global search used in a few well known optimization schemes under a single unified framework propped up by a sound probabilistic basis. In a way explained later, the ideas (or their possible generalizations) behind some of the existing optimization methods may sometimes be readily incorporated in the present setting by just tweaking the innovation process.
\par The rest of the paper is organized as follows. In Section 2, the problem of finding an extremum of an objective functional is posed as a martingale problem represented through an integro-differential equation, whose solution only realizes a local optimum. The integro-differential equation is discretized and weakly solved within an MC setting so as to circumvent its inherent circularity. Section 3 discusses several random perturbation schemes to arrive at the global optimum efficiently without getting stuck in local traps. By combining these tools in different ways, a few pseudo-codes are presented in Section 4. In Section 5, the performance of COMBEO is compared with DE and PSO in the context of several benchmark problems. In the same section, the new optimization approach is also applied for the quantitative recovery of boundary layer density variation in a high speed flow given the light travel time data from an experiment. Finally, the conclusions of this study are drawn in Section 6.

\section{Search for an extremum through a martingale problem} \label{martingale_problem}
Here it is demonstrated that functional extremization, subject to a generic set of equality constraints, can be posed as a martingale problem upon a proper characterization of the design variables within a probabilistic setting. However, before elaborating on this, a few fundamental features of the new evolutionary approach are listed below.
\par i)	At a given iterate, the solution is a random variable (taking values in the search space).
\par ii)	Hence the solution process along the iteration axis may be considered a stochastic process and the iterations must be so designed that the optimal solution is asymptotically approached by the mean (i.e. the first moment) of the solution process.
\par iii)	Since upon convergence the mean should be iteration-invariant, random fluctuations about the converged mean must assume the structure of a zero-mean stochastic process (along the iteration axis), thereby allowing us to identify the fluctuations as a noise process, e.g. a Brownian motion, or, more generally, a zero-mean martingale.

\par In posing the global optimization problem within a stochastic framework, a complete probability space $(\Omega ,\,\mathcal{F},\,P)$ \cite{15oksendal} is considered, within which the solution to the optimization problem must exist as an $\mathcal{F}$-measurable random variable. Here $\Omega $  , known as the population set (sample space), necessarily includes all possible candidate solutions from which a finite set of randomly chosen realizations or particles is evolved along the iteration axis $\tau $.  The introduction of $\tau$ as a positive monotonically increasing function on $\mathbb{R}$ is required to qualify the evolution of the solution as a continuously parametered stochastic process. A necessary aspect of the evolution is a random initial scatter imparted to the particles so as to search the sample space for solutions that extremize the cost functional whilst satisfying the posed constraints, if any. Since the extremal values, global or otherwise, of the objective functional may not be known {\it a priori}, the particles are updated iteratively (i.e. along $\tau $) by conditioning on the evolution history of a so called extremal cost process based on the available particles. The extremal cost process is defined as a stochastic process such that, for any $\xi  \in [0,\tau ]$, its mean is given by the available best objective functional based on its evaluations across the particles at the iteration step denoted by $\xi $. Now denoting by
$\mathcal{N}_\tau:= {\{ {{\mathcal{G}}_\xi }\} _{\xi  \in [0,\tau ]}}$ the filtration (i.e. statistical information pertaining to the evolution history) containing the increasing family of sub $\sigma$-algebras generated by the extremal cost process, the aim is to determine the update by conditioning the guessed or predicted solution on ${{\mathcal{N}}_\tau }$. The guessed solution could be a Brownian motion or a random walk along $\tau$. In case there are equality constraints, ${{\mathcal{N}}_\tau }$ also contains history for the best realized constraint until $\tau$. For a multivariate multimodal nonlinear objective functional $f(\mathbf{x}):{{\mathbb{R}}^n} \mapsto {\mathbb{R}}$ in ${\mathbf{x}} = \left\{ {{x^j}} \right\}_{j = 1}^n \in {{\mathbb{R}}^n}$, a point ${{\bf{x}}^ * }$ needs to be found such that $f({{\bf{x}}^ * }) \leqslant f({\bf{x}}),\,\forall \,{\bf{x}} \in {{\mathbb{R}}^n}$. Since the design variable ${\bf{x}}$ is evolved in $\tau$ as a stochastic process, it is parameterized as ${{\bf{x}}_\tau }: = {\bf{x}}(\tau )$. It may be noted that there may not be any inherent physical dynamics in its evolution, and in such cases ${{\bf{x}}_\tau }$ may be evolved as a zero-mean Brownian motion or random walk in $\Omega$ (even though, as will be shown later, this step is not absolutely necessary). Since only a finite number of iterations can be performed in practice, $\tau$ is discretized as ${\tau _0} < {\tau _1} < ... < {\tau _M}$ so that ${{\bf{x}}_\tau }$ is evolved over every $\tau$ increment. Thus, for the ${i^{{\text{th}}}}$ iteration with $\tau  \in ({\tau _{i - 1}},{\tau _i}]$, ${{\bf{x}}_\tau }$ may be thought of as being governed by the following stochastic differential equation (SDE):
\begin{equation}\label{process_eq}
d{{\bf{x}}_\tau } = d{{\bf{\xi }}_\tau }
\end{equation}
Here ${{\bf{\xi }}_\tau }$ is a zero-mean vector Brownian process with the covariance matrix ${\bf{g}}{{\bf{g}}^T} \in {{\mathbb{R}}^{n \times n}}$, where ${\bf{g}} \in {{\mathbb{R}}^{n \times n}}$ is the noise intensity matrix with its ${\left( {j,k} \right)^{{\text{th}}}}$ element denoted as ${g^{jk}}$. The discrete $\tau$-marching map for Eqn. (\ref{process_eq}) has the form:
\begin{equation}\label{process_EM}
{{\bf{x}}_i} = {{\bf{x}}_{i - 1}} + \Delta {{\bf{\xi }}_i}\,\,\,,\,\Delta {{\bf{\xi }}_i}: = {{\bf{\xi }}_i} - {{\bf{\xi }}_{i - 1}}
\end{equation}
where ${\left(  \cdot  \right)_i}$ stands for ${\left(  \cdot  \right)_{{\tau _i}}}$. Let the extremal cost process, generating the filtration ${{\mathcal{N}}_\tau }$, be denoted as ${\overset{\lower0.5em\hbox{$\smash{\scriptscriptstyle\frown}$}}{{f} _\tau }}$. Within a strictly deterministic setting, a smooth functional $f({\bf{x}})$ tends to become stationary, in the sense of a vanishing first variation, as ${\bf{x}}$ approaches an extremal value ${{\bf{x}}^ * }$. In the stochastic setting, a counterpart of this scenario could be that any future conditioning of the process ${{\bf{x}}_\tau }$ on ${{\mathcal{N}}_\xi }$, which is the filtration generated by ${\overset{\lower0.5em\hbox{$\smash{\scriptscriptstyle\frown}$}}{{f} _s}}$, $s < \xi$, till a past iteration (i.e. $\xi  < \tau$), identically yields the random variable ${{\bf{x}}_\xi }$ itself, i.e. $E({{\bf{x}}_\tau }|{{\mathcal{N}}_\xi }) = {{\bf{x}}_\xi }$, where $E$ represents the expectation operator. Via the Markov structure of ${{\bf{x}}_\tau }$ one may equivalently postulate that a necessary condition for the extremization of ${f_\tau }: = f({{\bf{x}}_\tau })$ is to require that $E({{\bf{x}}_\tau }|{\overset{\lower0.5em\hbox{$\smash{\scriptscriptstyle\frown}$}}{{f} _\xi} }) = {{\bf{x}}_\xi }$. This characterization renders the process ${{\bf{x}}_\tau }$ a martingale with respect to the extremal cost filtration ${\mathcal{N}_\tau}$, i.e. once extremized, any future conditional mean of ${{\bf{x}}_\tau }$ on the cost filtration remains iteration invariant. See \cite{18klebaner} for an introductory treatise on the theory of martingales. An equivalent way of stating this condition is based on an error or `innovation' process defined as ${\overset{\lower0.5em\hbox{$\smash{\scriptscriptstyle\frown}$}}{{f} _\tau }} - {f_\tau }$. An extremization would require updating ${{\bf{x}}_\tau }$ to statistically match the extremal process ${\bf{x}}_\tau ^*$ so that the innovation ${\overset{\lower0.5em\hbox{$\smash{\scriptscriptstyle\frown}$}}{{f} _\tau }} - f({\bf{x}}_\tau ^*)$ is driven to a zero mean martingale, e.g. a zero-mean Brownian motion or, more generally, a stochastic integral in Ito's sense (or, in a $\tau $-discrete setting, a zero-mean random walk). This enables one to pose the determination of an extremum as a martingale problem, originally conceptualized by Stroock and Varadhan \cite{16stroock_varadhan} to provide an improved setting for solutions to SDEs beyond Ito's theory. If the update scheme is effective, one anticipates ${{\bf{x}}_\tau }\to {\bf{x}}_\tau ^*$ asymptotically, i.e. as $\tau  \to \infty$. In other words, $E\left( {{\lim\limits_{\tau  \to \infty }}E({{\bf{x}}_\tau }|{{\mathcal{N}}_\tau })} \right) = E({\bf{x}}_\tau ^*) = {{\bf{x}}^ * }$ and, as the norm of the noise intensity associated with ${{\bf{x}}_\tau }$ approaches zero, ${\bf{x}}_\tau ^* \to {{\bf{x}}^*}$. However, since strictly zero noise intensity is infeasible within the stochastic setup, the solution in a deterministic setting may be thought of as a degenerate version (with a Dirac measure) of that obtained through the stochastic scheme.

\par Note that a ready extension of this approach is possible with multi-objective functions, which would merely require building up the innovation as a vector-valued process. Similarly, the approach may also be adapted for treating equality constraints, wherein zero-valued constraint functions, perturbed by zero-mean noise processes, may be used to construct the innovation processes. The easy adaptability of the current setup for multiple objective functions or constraints may be viewed as an advantage over most existing evolutionary optimization schemes, where a single cost functional needs to be constructed based on all the constraints, a feature that may possibly engender instability for a class of large dimensional problems.
\par Driving ${{\bf{x}}_\tau }$ to an  ${{\mathcal{N}}_\tau }$-martingale, or forcing the innovation process to a zero-mean martingale, may be readily achieved through a change of measures. In evolutionary algorithms, e.g. the GA, the accompanying change of measures requires the assignment of weights or fitness values to the current set of particles and a subsequent selection of those with higher fitness, e.g. by rejection sampling. For a better exploration of the sample space, the GA uses steps like `crossover' and `mutation'. These steps are biologically inspired safeguards against an intrinsic limitation of the weight-based update, which tends to diminish all the weights but one to zero, thereby leaving the realized particles nearly identical after a few iterations. This problem is referred to as that of weight collapse or particle impoverishment and precipitates premature convergence to a wrong solution. As a way out of this degeneracy in the realized population set (i.e. the ensemble), the present scheme employs a purely additive update, derived following a Girsanov change of measures $ P\to Q$ (see \cite{15oksendal} for a detailed exposition on the Girsanov transformation), which ensures that the innovation process, originally described under measure $P$, becomes a zero-mean martingale under the new measure $Q$. A basic tool in effecting this change of measure is the Radon-Nikodym derivative ${\Lambda _\tau }: = \frac{{dP}}
{{dQ}}$ (assuming absolute continuity of $Q$ w.r.t. $P$ and vice versa), a scalar valued random variable also called the likelihood ratio (or the fitness or the weight) that multiplicatively updates the design process ${{\bf{x}}_\tau}$ as ${{\bf{x}}_\tau }{\Lambda _\tau }$. In order to bypass the degeneracy problem noted above, one may obtain an additive update by expanding ${{\bf{x}}_\tau }{\Lambda _\tau }$ using Ito's formula \cite{15oksendal}. Within an MC set-up, an immediate consequence of the additive update is that particles with lower weights are never eliminated, but rather corrected to have more fitness by being driven closer to an extremum. Drawing analogy with Taylor's expansion of a smooth function(al) that obtains its first order term based on Newton's directional derivative, the present version of the additive correction may be interpreted as a non-Newton directional term that drives the innovation process to a zero-mean martingale and a precise form of this term is derived later in this section. In general, the innovation process may be a vector. One such case occurs when, by way of addressing possible numerical instability, a single cost functional is split into several so that each of the correspondent innovation is driven to a zero-mean martingale. This in turn ensures that the innovation corresponding to the original cost functional is also driven to a zero-mean martingale. For a simpler exposition of the basic ideas, the method below is presented using a single cost functional only. The formulation can be trivially extended for vector innovation processes of any finite dimension.

\par Within a $\tau$-discrete framework with $\tau  \in ({\tau _{i - 1}},{\tau _i}]$, ${{\bf{x}}_\tau }$ is evolved following Eqn. (\ref{process_eq}). An innovation constraint that may be satisfied for an accelerated (greedy) search in a neighborhood of an extremum is given by:
\begin{equation}\label{extremal_eq1}
{\overset{\lower0.5em\hbox{$\smash{\scriptscriptstyle\frown}$}}
 {{f} _\tau}} - f\left( {{\bf{x}_\tau }} \right) = \Delta {\eta _\tau }
\end{equation}
where ${\overset{\lower0.5em\hbox{$\smash{\scriptscriptstyle\frown}$}}{{f} _\tau }} \in {\mathbb{R}}$ is the extremal cost process, {\it{f}} the objective functional and $\Delta {\eta _\tau } = {\eta _\tau } - {\eta _{i - 1}} \in {\mathbb{R}}$ a  {\it{P}}-Brownian increment representing the diffusive fluctuations. Deriving the subsequent integro-differential equation for the search is best accomplished by recasting Eqn. (\ref{extremal_eq1}) as an SDE. Towards this, an ${{\mathcal{N}}_\tau }$-measureable process ${\overset{\lower0.5em\hbox{$\smash{\scriptscriptstyle\smile}$}}{{f} _\tau }}: = \overset{\lower0.5em\hbox{$\smash{\scriptscriptstyle\smile}$}}{{f}} \left( \tau  \right)$ may be constructed to arrive at the following incremental form:
\begin{equation}\label{extremal_eq2}
\Delta {\overset{\lower0.5em\hbox{$\smash{\scriptscriptstyle\smile}$}}{{f} _\tau}} : = {\overset{\lower0.5em\hbox{$\smash{\scriptscriptstyle\frown}$}}{{f} _\tau}}\Delta \tau  = f({{\bf{x}}_\tau })\Delta \tau  + \Delta {\eta _\tau }\Delta \tau;     \Delta \tau= \tau  - \tau _{i - 1}
\end{equation}
Here $\Delta {\tau _i}: = {\tau _i} - {\tau _{i - 1}}$ is taken as a `small' increment. Since the  $\tau$-axis is fictitious, $\Delta\tau$  is replaced by $\Delta {\tau _i} = {\tau _i} - {\tau _{i - 1}}$, so that Eqn. (\ref{extremal_eq2}) is modified to:

\begin{equation} \label{extremal_eq3}
\Delta {\overset{\lower0.5em\hbox{$\smash{\scriptscriptstyle\smile}$}}{{f} _\tau}} = f({{\bf{x}}_\tau })\Delta \tau  + \Delta {\eta _\tau }\Delta {\tau _i}
\end{equation}
which is essentially correspondent to the SDE:
\begin{equation} \label{extremal_eq4}
d{\overset{\lower0.5em\hbox{$\smash{\scriptscriptstyle\smile}$}}{{f} _\tau}} = f({{\bf{x}}_\tau })d\tau
+ \Delta {\tau _i}d{\eta_\tau }
\end{equation}

Note that the replacement of $\Delta \tau$ by $\Delta {\tau _i}$ merely modifies the intensity of the noise process $\Delta {\eta _\tau}$ in Eqn. (\ref{extremal_eq1}). The form of SDE (\ref{extremal_eq4}) has the desirable feature that the fictitious diffusion coefficient $\Delta {\tau _i}$ is an order `smaller' relative to the drift coefficient $f({{\bf{x}}_\tau })$. However, since ${\eta _\tau }$ is not standard Brownian, it is more convenient to rewrite Eqn. (\ref{extremal_eq4}) as:

\begin{equation} \label{extremal_eq5}
d{\overset{\lower0.5em\hbox{$\smash{\scriptscriptstyle\smile}$}}{{f} _\tau}} = f({{\bf{x}}_\tau })d\tau  + {\rho _\tau }d{W_\tau }
\end{equation}

${W_\tau}$ is standard {\it{P}}-Brownian and ${\rho _\tau}$ a more general form of (scalar-valued) noise intensity that may be made a function of $\tau$. For multi-objective optimization problems (where ${\overset{\lower0.5em\hbox{$\smash{\scriptscriptstyle\smile}$}}{{f} _\tau}}$ is a vector stochastic process), ${\rho _\tau}$ would be the intensity matrix and the expressions below are so written that they are valid for both scalar and vector cases. Eqn. (\ref{extremal_eq5}) may be rewritten as:

\begin{equation} \label{extremal_eq6}
d{\tilde f_\tau }: = \rho _\tau ^{ - 1}d{\overset{\lower0.5em\hbox{$\smash{\scriptscriptstyle\smile}$}}{{f} _\tau}} = h({{\bf{x}}_\tau })d\tau + d{W_\tau}
\end{equation}

where $h({{\bf{x}}_\tau }): = \rho _\tau ^{ - 1}f({{\bf{x}}_\tau })$. While it is possible, and even desirable, to replace the Brownian noise term above by one whose quadratic variation is zero (e.g. a Poisson martingale), such a modification is not central to the basic idea and will be considered in future extensions of the work. SDE (\ref{extremal_eq5}) is assumed to satisfy the standard conditions \cite{15oksendal} so as to ensure the existence of a weak solution. The ${{\mathcal{N}}_\tau }$-measurable locally optimal (extremal) solution may now be identified with the conditional mean $E\left( {{{\bf{x}}_\tau }|{{\mathcal{N}}_\tau }} \right)$. Considering a new measure {\it{Q}} under which ${{\bf{x}}_\tau}$ from Eqn. (\ref{process_eq}) satisfies Eqn. (\ref{extremal_eq5}), the conditional mean may be expressed via the generalized Bayes' formula as:

\begin{equation} \label{bayes_formula}
{\pi _\tau }\left( {\bf{x}} \right): = E\left( {{{\bf{x}}_\tau }|{{\mathcal{N}}_\tau }} \right)\, = \frac{{{E_Q}\left( {{{\bf{x}}_\tau }{\Lambda _\tau }|{{\mathcal{N}}_\tau }} \right)\,}}
{{{E_Q}\left( {{\Lambda _\tau }|{{\mathcal{N}}_\tau }} \right)\,}}
\end{equation}

where the expectation ${E_Q}\left(  \cdot  \right)$ is taken with respect to the new measure {\it{Q}} and ${\Lambda _\tau}$ is the scalar fitness given by:

\begin{equation} \label{girsanov}
{\Lambda _\tau } = \exp \left[ {\int_{{\tau _{i - 1}}}^\tau  {{h_s}d{{\tilde f}_s} - \frac{1}
{2}\int_{{\tau _{i - 1}}}^\tau  {h_s^T{h_s}ds} } } \right]
\end{equation}

As shown in the Appendix (a theorem and its corollary) using Ito's expansions of ${{\bf{x}}_\tau }{\Lambda _\tau }$ and ${({\Lambda _\tau })^{ - 1}}$, the incrementally additive updates to arrive at an extremum must follow from the differential equation:

\begin{equation} \label{ks_incr}
\begin{split}
d{\pi _\tau }\left( \bf{x} \right) = \left( {{\pi _\tau }\left( {\bf{x}}f \right) - {\pi _\tau }\left( \bf{x} \right){\pi _\tau }\left( f \right)} \right){\left( {{\rho _\tau }\rho _\tau ^T} \right)^{ - 1}}
\left( {d{{\overset{\lower0.5em\hbox{$\smash{\scriptscriptstyle\smile}$}}{{f}_\tau}} } - {\pi _\tau }\left( f \right)d\tau } \right)
\end{split}
\end{equation}

\begin{equation} \label{ks_int}
\begin{split}
{\pi _\tau }\left( \bf{x} \right) = {\pi _{i - 1}}\left( \bf{x} \right) +
\int_{\tau_{i - 1}}^\tau  {\left( {{\pi _s}\left( {\bf{x}}f \right) - {\pi _s}\left( \bf{x} \right){\pi _s}\left( f \right)} \right){{\left( {{\rho _s}\rho _s^T} \right)}^{ - 1}} \left( {d{{\overset{\lower0.5em\hbox{$\smash{\scriptscriptstyle\smile}$}}{{f}_s}}} - {\pi _s}\left( f \right)ds} \right)}
\end{split}
\end{equation}

Note that the directionality of a search process provided by the second (integral) term on the right hand side of the equation above may also be gauged from the fact the integrand in this term can be interpreted as a Malliavin-type derivative using Clarke-Ocone theorem \cite{clarke_ocone}. But the appearance of the unknown term ${\pi _\tau }\left( f \right)$ on the right hand side (RHS) prevents Eqn. (\ref{ks_incr}) or (\ref{ks_int}) to be qualified as an SDE in ${\pi _\tau }\left( \bf{x} \right)$. Indeed, a necessarily nonlinear dependence of {\it{f}} on ${{\bf{x}}_\tau }$ and the consequent non-Gaussianity of ${\pi _\tau }\left( f \right)$ would prevent writing the latter in terms of ${\pi _\tau }\left( {\bf{x}} \right)$. This results in the so called closure problem in solving for ${\pi _\tau }\left( \bf{x} \right)$.
\par While a direct solution of Eqn. (\ref{ks_incr}) or (\ref{ks_int}) yields an extremum (i.e. a local extremum only) in principle, exact/analytical solutions are infeasible owing to the circularity inherent in the closure problem. This has a parallel in nonlinear stochastic filtering, wherein the Kushner-Stratonovich equation \cite{19kushner} (an equivalent of Eqn. (\ref{ks_incr})) also suffers from a similar circularity problem. Motivated by the MC filters used to solve nonlinear filtering problems \cite{20sarkar_physicaD, 21EnKS}, an MC scheme may be developed for a numerical treatment of Eqn. (\ref{ks_incr}) or (\ref{ks_int}). A two stage strategy, viz. prediction and update, may apparently be considered, even though, as we will soon see, the prediction step could be entirely eliminated in the final scheme. The prediction step, as in most evolutionary optimization schemes, aims at an initial random exploration based on Eqn. (\ref{process_eq}). Consider the ${i^{{\text{th}}}}$ iteration, i.e. $\tau  \in ({\tau _{i - 1}},{\tau _i}]$ and let {\it{N}} denote the ensemble size so that one realizes {\it{N}} predicted particles or MC candidates, $\left\{ {{\bf{x}}_\tau ^{\left( j \right)}} \right\}_{j = 1}^N$ that must be updated via Eqn.  (\ref{ks_int}). For an MC-based numerical solution to Eqn.  (\ref{ks_int}), a sample-averaged form of the equation is first written as:

\begin{equation} \label{ks_ensemble_approx}
\begin{split}
\pi _\tau ^N\left( \bf{x} \right) = \pi _{i - 1}^N\left( \bf{x} \right) +
 \int_{{\tau _{i - 1}}}^\tau  {\left( {\pi _s^N\left( {\bf{x}}f \right) - \pi _s^N\left( \bf{x} \right)\pi _s^N\left( f \right)} \right){{\left( {{\rho _s}\rho _s^T} \right)}^{ - 1}}}
 \left( {d{{\overset{\lower0.5em\hbox{$\smash{\scriptscriptstyle\smile}$}}{{f}_s}}} - \pi _s^N\left( f \right)ds} \right)
\end{split}
\end{equation}

${\pi ^N}\left( . \right) = (1/N)\sum\limits_{j = 1}^N {{{\left( . \right)}^{\left( j \right)}}} $ is the ensemble-averaged approximation to the conditional mean $\pi \left( . \right)$. A particle-wise representation of Eqn. (\ref{ks_ensemble_approx}) is given by:
\begin{equation} \label{ks_particle}
\begin{split}
{\bf{X}_\tau } = {{\bf{X}}_{i - 1}} +
 \frac{1}
{N}\int_{{\tau _{i - 1}}}^\tau  {\left\{ {{\bf{X}_s}{{\bf{F}}_s}^T - {{\hat {\bf{X}}}_s}{{\hat {\bf{F}}}_s}^T} \right\}{{\left( {{\rho _s}\rho _s^T} \right)}^{ - 1}}\left\{ {d{{\overset{\lower0.5em\hbox{$\smash{\scriptscriptstyle\smile}$}}{{\bf{F}}_s}}} - {{\bf{F}}_s}ds} \right\}}
\end{split}
\end{equation}
where ${\bf{X}_\tau }: = [{\bf{x}}_\tau ^{(1)},{\bf{x}}_\tau ^{(2)},...{\bf{x}}_\tau ^{(N)}]$, ${{\bf{F}}_\tau }: = [f_\tau ^{(1)},f_\tau ^{(2)},...,f_\tau ^{(N)}]$, ${\hat {\bf{X}}_\tau }: = \pi _\tau ^N\left( {\bf{x}} \right){\bf{r}} \in {{\mathbb{R}}^{n \times N}}$,
${\hat {\bf{F}}_\tau }: = \pi _\tau ^N\left( f \right){\bf{r}} \in {{\mathbb{R}}^N}$ and $d{\overset{\lower0.5em\hbox{$\smash{\scriptscriptstyle\smile}$}}{{{\bf{F}}} _\tau}}: = d{\overset{\lower0.5em\hbox{$\smash{\scriptscriptstyle\smile}$}}{{f} _\tau}}{\bf{r}} \in {{\mathbb{R}}^N}$. ${\bf{r}} = \left\{ {1,\,1,\,...,\,1} \right\} \in {{\mathbb{R}}^N}$ is an {\it N}-dimensional row vector with entries 1. Note that the second term on the RHS of Eqn. (\ref{ks_particle}) is the additive update/correction term. For solving Eqn. (\ref{ks_particle}), the MC approximation to Eqn. (\ref{ks_int}), a $\tau$-discrete numerical scheme is required. Such a scheme would typically involve the following two steps.
\par \text{}
\par {\it{a) Prediction}}:
\par \text{}
\par The predicted particle set ${\tilde {\bf{X}}_\tau } = [\tilde {\bf{x}}_\tau ^{(1)},...,\tilde {\bf{x}}_\tau ^{(N)}]$ at $\tau$  is generated using an Euler-Maruyama (EM) discretized map:

\begin{equation} \label{EM-prediction}
{\tilde {\bf{X}}_\tau } = {{\bf{X}}_{i - 1}} + \Delta {{\bf{\Psi}} _\tau}
\end{equation}
where $\Delta {{\bf{\Psi}} _\tau }: = [\Delta {\bf{\xi }}_\tau ^{\left( 1 \right)},\,...,\,\Delta {\bf{\xi }}_\tau ^{\left( N \right)}]$, $\Delta {\bf{\xi }}_\tau ^{\left( 1 \right)} = {\bf{\xi }}_\tau ^{\left( 1 \right)} - {\bf{\xi }}_{i - 1}^{\left( 1 \right)}$ etc.
\par \text{}
\par {\it{b) Additive Update}}:
\par \text{}
\par The update to the predicted particles is through an EM approximation to the integral in Eqn. (\ref{ks_particle}). Higher order integration schemes could also be considered, especially for evaluating the correction integral. The discrete update equation is presently given by:

\begin{equation} \label{ks_particle2}
\begin{split}
{{\bf{X}}_\tau } = {\tilde {\bf{X}}_\tau } +
\frac{1}
{N}\left\{ {{{\tilde {\bf{X}}}_\tau }{{\tilde {\bf{F}}}_\tau }^T - {{\tilde {\hat {\bf{X}}}}_\tau }{{\tilde{\hat {\bf{F}}} }_\tau }^T} \right\}{\left( {{\rho _\tau }\rho _\tau ^T} \right)^{ - 1}}\left\{ {\Delta {{\overset{\lower0.5em\hbox{$\smash{\scriptscriptstyle\smile}$}}{{{\bf{F}}}_\tau} } } - {{\tilde {\bf{F}}}_\tau }\Delta \tau } \right\}
\end{split}
\end{equation}
where $\Delta {\overset{\lower0.5em\hbox{$\smash{\scriptscriptstyle\smile}$}}{{{\bf{F}}} _\tau}} = \Delta {\overset{\lower0.5em\hbox{$\smash{\scriptscriptstyle\smile}$}}{{f} _\tau}}{\bf{r}}$, ${\tilde {\bf{F}}_\tau }: = [\tilde f_\tau ^{(1)},...,\tilde f_\tau ^{(N)}]$ and the predicted solution, ${\tilde {\bf{x}}_\tau }$, is used to compute ${\tilde f_\tau }: = f\left( {{{\tilde {\bf{x}}}_\tau }} \right)$. Moreover, ${\tilde {\hat {\bf{F}}}_\tau }: = \pi _\tau ^N\left( {\tilde f} \right)\bf{r}$. Eqn. (\ref{ks_particle2}) may be recast as:

\begin{equation} \label{ks_particle3}
\begin{split}
{{\bf{X}}_\tau } = {\tilde {\bf{X}}_\tau } + \frac{1}
{N}\left\{ {\left( {{{\tilde {\bf{X}}}_\tau } - {{\tilde {\hat {\bf{X}}}}_\tau }} \right){{\tilde {\bf{F}}}_\tau }^T + {{\tilde \hat {\bf{X}}}_\tau }{{\left( {{{\tilde {\bf{F}}}_\tau } - {{\tilde {\hat {\bf{F}}}}_\tau }} \right)}^T}} \right\}
{\left( {{\rho _\tau }\rho _\tau ^T} \right)^{ - 1}}\left\{ {\Delta {{\overset{\lower0.5em\hbox{$\smash{\scriptscriptstyle\smile}$}}{{{\bf{F}}}_\tau} } } - {{\tilde {\bf{F}}}_\tau }\Delta \tau } \right\}
\end{split}
\end{equation}
Recall from Eqn. (\ref{extremal_eq2}) that $\Delta {\overset{\lower0.5em\hbox{$\smash{\scriptscriptstyle\smile}$}}{{f} _\tau} }: = {\overset{\lower0.5em\hbox{$\smash{\scriptscriptstyle\frown}$}}{{f} _\tau} }\Delta \tau $  so that Eqn. (\ref{ks_particle3}) may be rearranged as:

\begin{equation} \label{ks_particle4}
\begin{split}
{{\bf{X}}_\tau } = {\tilde {\bf{X}}_\tau } + \frac{1}
{N}\left\{ {\left( {{{\tilde {\bf{X}}}_\tau } - {{\tilde {\hat {\bf{X}}}}_\tau }} \right)\left( {{{\tilde {\bf{F}}}_\tau }^T\Delta \tau } \right) + \left( {{{\tilde {\hat {\bf{X}}}}_\tau }\Delta \tau } \right){{\left( {{{\tilde {\bf{F}}}_\tau } - {{\tilde {\hat {\bf{F}}}}_\tau }} \right)}^T}} \right\}{\left( {{\rho _\tau }\rho _\tau ^T} \right)^{ - 1}}\left\{ {{{\overset{\lower0.5em\hbox{$\smash{\scriptscriptstyle\frown}$}}{{{\bf{F}}}_\tau} } } - {{\tilde {\bf{F}}}_\tau }} \right\}
\end{split}
\end{equation}

where ${{\bf{\overset{\lower0.5em\hbox{$\smash{\scriptscriptstyle\frown}$}}{{{\bf{F}}}_\tau} }} } = {\overset{\lower0.5em\hbox{$\smash{\scriptscriptstyle\frown}$}}{{f} _\tau} }{\bf{r}} \in {{\mathbb{R}}^N}$. When the particles are somewhat away from a local extremum (e.g. during the initial stages of evolution), the (norm of the) correction term is large. Hence the particles traverse more in the search space. In such cases, the innovation process would not behave as a zero-mean martingale as it would have a significant drift component. Since evolutions in this regime may have sharper  $\tau$-gradients, it is appropriate to modify the coefficient matrix in Eqn. 3.6 so as to incorporate information on these gradients through previous estimates. Thus ${\tilde {\bf{F}}_\tau }^T\Delta \tau $ and ${\tilde{\hat{\bf{X}}}}_\tau \Delta \tau$ are replaced respectively by the following approximations:

\begin{equation} \label{ito_expand1}
{\tilde {\bf{F}}_\tau }^T\Delta \tau  \approx \left( {{{\tilde {\bf{F}}}_\tau }^T\tau  - {{\tilde {\hat {\bf{F}}}}_{i - 1}}^T{\tau _{i - 1}} - {\Delta {\tilde {\hat {\bf{F}}}}_\tau }^T\tau } \right)
\end{equation}

\begin{equation} \label{ito_expand2}
{\tilde {\hat {\bf{X}}}}_\tau \Delta \tau  \approx \left( {{\tilde {\hat {\bf{X}}}}_\tau }\tau  - {{\tilde {\hat {\bf{X}}}}_{i - 1}{\tau _{i - 1}}} \right)
\end{equation}

Note that we have used Ito's formula while approximating the RHS in Eqn. (\ref{ito_expand1}).  Using Eqn. (\ref{ito_expand1}) and (\ref{ito_expand2}), Eqn. (\ref{ks_particle4}) may be modified as:
\begin{equation} \label{update1}
\begin{split}
{{\bf{X}}_\tau }={\tilde {\bf{X}}_\tau }+ \frac{1}{N}{\left\{\begin{array}{c}\left( {{{\tilde {\bf{X}}}_\tau } - {{\tilde {\hat {\bf{X}}}}_\tau }} \right)\left( {{{\tilde {\bf{F}}}_\tau }^T\tau  - {{\tilde {\hat {\bf{F}}}}_{i - 1} }^T{\tau _{i - 1}} - \Delta {{\tilde {\hat {\bf{F}}}}_\tau }^T\tau } \right)\\+ \left( {{{\tilde {\hat {\bf{X}}}}_\tau }\tau  - {{\tilde {\hat {\bf{X}}}}_{i - 1}}{\tau _{i - 1}}} \right){\left( {{{\tilde {\bf{F}}}_\tau } - {{\tilde \hat {\bf{F}}}_\tau }} \right)^T}\end{array}\right\}}{\left( {{\rho _\tau }\rho _\tau ^T} \right)^{ - 1}}\left\{ {{{\overset{\lower0.5em\hbox{$\smash{\scriptscriptstyle\frown}$}}
 {{\bf{F}}} }_\tau } - {{\tilde {\bf{F}}}_\tau }} \right\}
\end{split}
\end{equation}

Once the converged estimate is obtained, the innovation noise covariance ${\rho _\tau }\rho _\tau ^T$ should satisfy the identity:
\begin{equation} \label{cov}
\begin{split}
{\rho _\tau }\rho _\tau ^T \approx \pi _\tau ^N\left( {\left( {{{\overset{\lower0.5em\hbox{$\smash{\scriptscriptstyle\frown}$}}{{f}_\tau} } } - {{\tilde f}_\tau }} \right){{\left( {{{\overset{\lower0.5em\hbox{$\smash{\scriptscriptstyle\frown}$}}{{f}_\tau} } } - {{\tilde f}_\tau }} \right)}^T}} \right)
\end{split}
\end{equation}
$$= \frac{1}
{{N - 1}}\left( {\left( {{{\overset{\lower0.5em\hbox{$\smash{\scriptscriptstyle\frown}$}}{{{\bf{F}}}_\tau} } } - {{\tilde {\bf{F}}}_\tau }} \right) - \left( {{{\overset{\lower0.5em\hbox{$\smash{\scriptscriptstyle\frown}$}}{{{\bf{F}}}_\tau} } } - {{\tilde {\hat {\bf{F}}}}_\tau }} \right)} \right){\left( {\left( {{{\overset{\lower0.5em\hbox{$\smash{\scriptscriptstyle\frown}$}}{{{\bf{F}}}_\tau} } } - {{\tilde {\bf{F}}}_\tau }} \right) - \left( {{{\overset{\lower0.5em\hbox{$\smash{\scriptscriptstyle\frown}$}}{{{\bf{F}}}_\tau} } } - {{\tilde {\hat {\bf{F}}}}_\tau }} \right)} \right)^T}
$$

Prior to convergence to an extremum, the (norm of the) RHS of the equation above would typically be relatively large. Thus, one could impart higher diffusion to the search in the initial stages by replacing ${\rho _\tau }\rho _\tau ^T$ in Eqn. (\ref{update1}) by:
\begin{equation} \label{cov1}
\begin{split}
\alpha \frac{1}
{{N - 1}}\left( {\left( {{{\overset{\lower0.5em\hbox{$\smash{\scriptscriptstyle\frown}$}}{{{\bf{F}}}_\tau} } } - {{\tilde {\bf{F}}}_\tau }} \right) - \left( {{{\overset{\lower0.5em\hbox{$\smash{\scriptscriptstyle\frown}$}}{{{\bf{F}}}_\tau} } } - {{\tilde \hat {\bf{F}}}_\tau }} \right)} \right){\left( {\left( {{{\overset{\lower0.5em\hbox{$\smash{\scriptscriptstyle\frown}$}}{{{\bf{F}}}_\tau} } } - {{\tilde {\bf{F}}}_\tau }} \right) - \left( {{{\overset{\lower0.5em\hbox{$\smash{\scriptscriptstyle\frown}$}}{{{\bf{F}}}_\tau} } } - {{\tilde \hat {\bf{F}}}_\tau }} \right)} \right)^T} + \left( {1 - \alpha } \right){\rho _\tau }\rho _\tau ^T
 \end{split}
 \end{equation}

 Here  $\alpha\in(0,1]$. It is typically taken as 0.8 in the numerical illustrations based on {\emph{pseudo-codes 1, 2 } and {\emph{3}}. Eqn. (\ref{update1}) thus takes the final form:
\begin{equation} \label{update3}
\begin{split}
{{{\bf{X}}}_{\tau }}={{\tilde{{\bf{X}}}}_{\tau }}+\frac{1}{N}{\left\{\begin{array}{c}{\left( {{{\tilde {\bf{X}}}_\tau } - {{\tilde{ \hat {\bf{X}}}}_\tau }} \right)\left( {{{\tilde {\bf{F}}}_\tau }^T\tau  - {{\tilde {\hat {\bf{F}}}}_{i - 1}}^T{\tau _{i - 1}} - \Delta {{\tilde {\hat {\bf{F}}}}_\tau }^T\tau } \right)}\\{ + \left( {{{\tilde {\hat {\bf{X}}}}_\tau }\tau  - {{\tilde {\hat {\bf{X}}}}_{i - 1}}{\tau _{i - 1}}} \right){{\left( {{{\tilde {\bf{F}}}_\tau } - {{\tilde {\hat {\bf{F}}}}_\tau }} \right)}^T}}\end{array}\right\}}.\\\left\{\begin{array}{c}{\alpha \frac{1}
{{N - 1}}\left( {\left( {{{\overset{\lower0.5em\hbox{$\smash{\scriptscriptstyle\frown}$}} {{{\bf{F}}}_\tau} } } - {{\tilde {\bf{F}}}_\tau }} \right) - \left( {{{\overset{\lower0.5em\hbox{$\smash{\scriptscriptstyle\frown}$}} {{{\bf{F}}}_\tau} } } - {{\tilde {\hat {\bf{F}}}}_\tau }} \right)} \right)}.
\\{{{\left( {\left( {{{\overset{\lower0.5em\hbox{$\smash{\scriptscriptstyle\frown}$}}{{{\bf{F}}}_\tau} } } - {{\tilde {\bf{F}}}_\tau }} \right) - \left( {{{\overset{\lower0.5em\hbox{$\smash{\scriptscriptstyle\frown}$}}{{{\bf{F}}}_\tau} } } - {{\tilde {\hat {\bf{F}}}}_\tau }} \right)} \right)}^T} + \left( {1 - \alpha } \right){\rho _\tau }\rho _\tau ^T}
\end{array}\right\}^{-1}\left\{ {{{\overset{\lower0.5em\hbox{$\smash{\scriptscriptstyle\frown}$}}{{{\bf{F}}}_\tau} } } - {{\tilde {\bf{F}}}_\tau }} \right\}
\end{split}
\end{equation}

A more concise form of the update equation is:
\begin{equation} \label{update2}
{{\bf{X}}_\tau } = {\tilde {\bf{X}}_\tau } + {{\bf{\tilde {\bf{G}}}}_\tau }\left\{ {{{\overset{\lower0.5em\hbox{$\smash{\scriptscriptstyle\frown}$}}{{{\bf{F}}}_\tau} } } - {{\tilde {\bf{F}}}_\tau }} \right\}
\end{equation}
where the gain-like update coefficient matrix is given by:

$${{{\bf{\tilde G}}}_\tau }: = \frac{1}
{N}\left\{\begin{array}{c}{\left( {{{\tilde {\bf{X}}}_\tau } - {{\tilde {\hat {\bf{X}}}}_\tau }} \right)\left( {{{\tilde {\bf{F}}}_\tau }^T\tau  - {{\tilde {\hat {\bf{F}}}}_{i - 1}}^T{\tau _{i - 1}} - \Delta {{\tilde {\hat {\bf{F}}}}_\tau }^T\tau } \right)}\\{ + \left( {{{\tilde {\hat {\bf{X}}}}_\tau }\tau  - {{\tilde {\hat {\bf{X}}}}_{i - 1}}{\tau _{i - 1}}} \right){{\left( {{{\tilde {\bf{F}}}_\tau } - {{\tilde {\hat {\bf{F}}}}_\tau }} \right)}^T}}\end{array}\right\}$$
$$\left\{\begin{array}{c}{\alpha \frac{1}
{{N - 1}}\left( {\left( {{{\overset{\lower0.5em\hbox{$\smash{\scriptscriptstyle\frown}$}}{{{\bf{F}}}_\tau} } } - {{\tilde {\bf{F}}}_\tau }} \right) - \left( {{{\overset{\lower0.5em\hbox{$\smash{\scriptscriptstyle\frown}$}}{{{\bf{F}}}_\tau} } } - {{\tilde {\hat {\bf{F}}}}_\tau }} \right)} \right)}{{{\left( {\left( {{{\overset{\lower0.5em\hbox{$\smash{\scriptscriptstyle\frown}$}}{{{\bf{F}}}_\tau} } } - {{\tilde {\bf{F}}}_\tau }} \right) - \left( {{{\overset{\lower0.5em\hbox{$\smash{\scriptscriptstyle\frown}$}}{{{\bf{F}}}_\tau} } } - {{\tilde {\hat {\bf{F}}}}_\tau }} \right)} \right)}^T{ + \left( {1 - \alpha } \right){\rho _\tau }\rho _\tau ^T}}}\end{array}\right\}^{-1}$$

In the update strategy of Eqn. (\ref{update2}), one may still improve on the search space exploration my multiplying the gain-weighted innovation term by a scalar factor ${\beta _\tau } < 1$. This is equivalent to increasing the noise covariance and hence allowing the particles to be more diffusive or `explorative'. Thus the update equation becomes:
\begin{equation} \label{update_final}
{{\bf{X}}_\tau } = {\tilde {\bf{X}}_\tau } + {\beta _\tau }{{\bf{\tilde G}}_\tau }\left\{ {{{\overset{\lower0.5em\hbox{$\smash{\scriptscriptstyle\frown}$}}{{{\bf{F}}}_\tau} } } - {{\tilde {\bf{F}}}_\tau }} \right\}
\end{equation}

\section{Coalescence, scrambling and relaxation: schemes for global search} \label{coalescence_scrambling}
The possible trapping of particles in local extrema is a major challenge to any global optimization scheme. For ${\beta _\tau } \geqslant 1$, the scheme described in Eqn. (\ref{update_final}) yields a greedy search that may often end up in a local extremum. Although it seems possible to avoid the local traps through the innovation noise, whose intensity could be tuned by the scalar factor ${\beta _\tau }$, such an approach may not be quite effective. As a local extremum is approached, the `strength' (or norm) of the update term would be small and consequently its sensitivity to variations in ${\beta _\tau }$ would also be poor. This makes the choice of ${\beta _\tau }$ difficult (e.g. necessitating $\beta_\tau$  to be too small for the search scheme to be efficient) and less effective for the global search. Moreover, smaller ${\beta _\tau }$  also implies larger diffusion and hence poorer convergence. A more effective way out is a random perturbation applied to the particles so as to force out the ones trapped in the local wells. A general perturbation scheme, combining three basic approaches referred to as `coalescence', `scrambling' and `relaxation', is now described. While the coalescence component is implemented through yet another martingale problem, the scrambling part requires a perturbation kernel. The relaxation component, on the other hand, requires accepting improvements with some probability. The importance of the random perturbation steps, in the present context, may also be gauged from the following fact. For a class of problems considered later, the faster convergence to a local extremum, engendered by the greedy scheme (\ref{update_final}), may force wrong convergence despite the application of random perturbations in conjunction with the greedy local search. Hence it would be worth exploring alternative search schemes, which completely eliminate scheme (\ref{update_final}) and rely on the random perturbation steps alone.
\par Since the aim is to pose `coalescence' as a martingale problem, the associated update should share the same generic structure as Eqn. (\ref{update_final}), which is recast below for the ${\it{j}}^{th}$ candidate:

\begin{equation} \label{update_3}
{\bf{x}}_i^{(j)} = {\bf{\tilde x}}_i^{(j)} + {\bf{\tilde C}}_i^{\left( j \right)}
\end{equation}

where ${\bf{\tilde C}}_i^{\left( j \right)}: = {\beta _i}{{\bf{\tilde G}}_i}\left\{ {{{\overset{\lower0.5em\hbox{$\smash{\scriptscriptstyle\frown}$}}{{f}_i}}} - \tilde f_i^{\left( j \right)}} \right\}$. The basic idea here is to provide to the evolving solution layers of random perturbation, whose inverse intensity may be formally indexed by a positive integer  $l$ such that the perturbation vanishes as $l \to \infty $. Within the  $\tau$-discrete setting, we start with the prediction ${\tilde {\bf{x}}_i}$ and denote by ${}^l\Delta {\hat {\bf{x}}_i} = {}^l{{\bf{x}}_i} - {}^l{\tilde {\bf{x}}_i}$ the randomly perturbed, ${\it{l}}$-indexed increment so that ${}^\infty \Delta {\hat {\bf{x}}_i} \to \Delta {{\bf{x}}_i}$ as the random perturbations vanish asymptotically. During the $i^{th}$ iteration, the perturbed increment is arrived at using two transitional increments, ${}^l{{\bf{u}}_i}$ and ${}^l{{\bf{v}}_i}$ determined by two perturbation operators, say ${\bf{T1}}$ and ${\bf{T2}}$ respectively. While the operator ${\bf{T1}}$ corresponds to the `local search' and/or `coalescence' operations, ${\bf{T2}}$ encapsulates the `scrambling' and `relaxation' operations. The transitions may be indicated as ${}^l\Delta {{\bf{x}}_i}\mathop  \to \limits_{{\bf{T1}}} {}^l{{\bf{u}}_i}\mathop  \to \limits_{{\bf{T2}}} {}^l{{\bf{v}}_i}\mathop  \to \limits_{{\bf{T3}}} {}^l\Delta {\hat {\bf{x}}_i}$ where ${\bf{T3}}$ is a selection operator, commonly used with most evolutionary optimization schemes in some form or the other. Here ${}^l\Delta {\hat {\bf{x}}_i}$ is the final increment at ${\tau _i}$, i.e. ${}^l{{\bf{x}}_i}: = {}^l{\tilde {\bf{x}}_i} + {}^l\Delta {\hat {\bf{x}}_i}$ , which is input to the next iteration. ${}^l\Delta {{\bf{x}}_i} = {}^l{\tilde {\bf{x}}_i} - {}^l{{\bf{x}}_{i - 1}}$ denotes the predicted increment. Ideally, one may start the iterations with small ${\it{l}}$ (i.e. high perturbation intensity), which is then gradually increased with progressing iterations. However, in the current implementations of COMBEO, the perturbation intensity is kept uniform all through the iterations. This is possible as, upon convergence to the global optimum, the applied perturbations merely yield zero-mean random fluctuations about the optimum, which are averaged out when the sample expectation operation is performed in the MC simulation. In view of this and for notational ease, the left superscript ${\it{l}}$ is removed from the variables in the discussion to follow and the perturbed nature of the variables should be clear from the context. The operators are now defined below.
\par {\bf{Operator T1: Local search and coalescence}}
\par The operation for local search has been described in Section II and through Eqn. (\ref{update_3}). The operation of `coalescence' is now outlined. This perturbation is motivated by the observation that the probability density function (PDF, if it exists) associated with the converged measure ${\pi _\tau }(.)$ should be unimodal, with its only peak located at the global extremum. Thus, when convergence to the global extremum occurs, all the particles should coalesce at the optimum point, except for a zero-mean noisy scatter around the latter. Ideally, for the sake of optimization accuracy, the noisy scatter should also have a low intensity so as to keep sample fluctuations under control. Once the global optimization scheme converges, the noisy scatter should then behave as a zero-mean martingale as a function of $\tau$ and with a unimodal transitional PDF. A zero-mean Brownian motion, which has a Gaussian PDF, is one such martingale. Clearly, such a property does not hold away from the global optimum, where the PDF would be multi-modal with a peak located at every local extremum detected by the algorithm.
\par Now, the aim is to obtain a scheme (or rather a family of schemes) to force the coalescence of particles. Consider the update of the ${j}^{th}$ particle ${\bf{x}}_\tau ^{(j)}$ such that coalescence of particles can be enforced. The random scatter around ${\bf{x}}_\tau ^{(j)}$ could be quantified by ${\delta _\tau }(j) = {\bf{x}}_\tau ^{({{\bf{\sigma }}_1}\left( j \right))} - {\bf{x}}_\tau ^{(j)}$, where  ${{\bf{\sigma }}_1}\left( j \right)$ denotes a random permutation on the indexing set $\{ 1,N\} \backslash \{ j\}$ based on a uniform measure. The goal is then to drive ${\delta _\tau }(j) = {\bf{x}}_\tau ^{({{\bf{\sigma }}_1}\left( j \right))} - {\bf{x}}_\tau ^{(j)}$ to a zero-mean vector Brownian increment $\Delta \eta _\tau ^c$ with intensity matrix $\rho _\tau ^c$  (typically assumed to be diagonal with the entries chosen uniformly for all ${\it{j}}$). In this case, `coalescence' then implies that, in the limit of the noise intensity in ${\delta _\tau }(j)$ approaching zero, all the particles would tend to coalesce into a single location at the global extremum. Thus, similar to the innovation ${\overset{\lower0.5em\hbox{$\smash{\scriptscriptstyle\frown}$}}{{f} _\tau} } - f\left( {{{\mathbf x}_\tau }} \right)$ on the left hand side (LHS) of Eqn. (\ref{extremal_eq1}), one treats ${\bf{x}}_\tau ^{({{\bf{\sigma }}_1}\left( j \right))} - {\bf{x}}_\tau ^{(j)}$ as yet another innovation process so that, upon convergence, the identity ${\bf{x}}_\tau ^{({{\bf{\sigma }}_1}\left( j \right))} - {\bf{x}}_\tau ^{(j)} = \Delta \eta _\tau ^c$ holds almost surely. Here $\Delta \eta _\tau ^c$ is responsible for an additional layer of randomness that gives every particle ${\bf{x}}_\tau ^{(j)}$ the structure of a stochastic process. Accordingly, the extremal cost filtration ${{\mathcal{N}}_\tau }$ must be suitably expanded/modified to include the sub-filtration generated by $\Delta \eta _s^c$ for $s \leqslant \tau $. If the coalescence innovation, in the form as indicated above, is included within our search process, Eqn. (\ref{update_3}) must be modified as:
\begin{equation} \label{update4}
\begin{split}
{\bf{x}}_i^{(j)} = {\bf{\tilde x}}_i^{(j)} + {\bf{\tilde D}}_i^{\left( j \right)}  \\
{\text{or }} {\bf{u}}_i^{(j)} = {\bf{\tilde D}}_i^{\left( j \right)}
\end{split}
\end{equation}
where
\begin{equation} \label{correction_term}
\begin{split}
{\bf{\tilde D}}_i^{\left( j \right)}: = {\beta _i}{{\bf{\tilde G}}_i}{\bf{\tilde I}}_i^{\left( j \right)} \\
{\bf{\tilde I}}_i^{\left( j \right)}: = \left\{\begin{array}{c}{{{\overset{\lower0.5em\hbox{$\smash{\scriptscriptstyle\frown}$}}{{f}_i} }} - \tilde f_i^{\left( j \right)}}\\{{\bf{\tilde x}}_i^{\left( {{{\bf{\sigma }}_1}\left( j \right)} \right)} - {\bf{\tilde x}}_i^{\left( j \right)}}\end{array}\right\}
\end{split}
\end{equation}
Recall that over-tildes indicate either the predicted particles or functions evaluated using the predicted particles, as appropriate. With a convenient abuse, the same notation for the gain-like update coefficient matrix ${{\bf{\tilde G}}_i}$ is retained in Eqn. (\ref{correction_term}) (used earlier in the local update Eqn. (\ref{update2})). ${{\bf{\tilde G}}_i}$ may now be computed as:

\begin{equation} \label{gain_matrix}
\begin{split}
{{{\bf{\tilde G}}}_i}: = \frac{1}{N}\left\{\begin{array}{c}{\left( {{{\tilde {\bf{X}}}_i} - {{\tilde {\hat {\bf{X}}}}_i}} \right)\left( {{{\tilde {\bf{F}}}_i}^T{\tau _i} - {{\tilde {\hat {\bf{F}}}}_{i - 1}}^T{\tau _{i - 1}} - \Delta {{\tilde {\hat {\bf{F}}}}_i}^T{\tau _i}} \right)}\\{ + \left( {{{\tilde {\hat {\bf{X}}}}_i}{\tau _i} - {{\tilde {\hat {\bf{X}}}}_{i - 1}}{\tau _{i - 1}}} \right){{\left( {{{\tilde {\bf{F}}}_i} - {{\tilde {\hat {\bf{F}}}}_i}} \right)}^T}}\end{array}\right\}.\\
\left\{\begin{array}{c}{\alpha \frac{1}
{{N - 1}}\left( {\left( {{{\overset{\lower0.5em\hbox{$\smash{\scriptscriptstyle\frown}$}}{{{\bf{F}}}_i} }} - {{\tilde {\bf{F}}}_i}} \right) - \left( {{{\overset{\lower0.5em\hbox{$\smash{\scriptscriptstyle\frown}$}}{{{\bf{F}}}_i} }} - {{\tilde {\hat {\bf{F}}}}_i}} \right)} \right)}.\\{{{\left( {\left( {{{\overset{\lower0.5em\hbox{$\smash{\scriptscriptstyle\frown}$}}{{{\bf{F}}} }_i}} - {{\tilde {\bf{F}}}_i}} \right) - \left( {{{\overset{\lower0.5em\hbox{$\smash{\scriptscriptstyle\frown}$}}{{{\bf{F}}}_i} }} - {{\tilde {\hat {\bf{F}}}}_i}} \right)} \right)}^T} + \left( {1 - \alpha } \right){{\bf{\gamma }}_i}{\bf{\gamma }}_i^T}\end{array}\right\}^{-1}
\end{split}
\end{equation}

\begin{equation} \label{cov_matrix}
{{\mathbf{\gamma }}_{i}}\mathbf{\gamma }_{i}^{T}:=\left[ \begin{matrix}
   {{\rho }_{i}}\rho _{i}^{T} & \mathbf{0}  \\
   \mathbf{0} & \rho _{i}^{c}{{(\rho _{i}^{c})}^{T}}  \\
\end{matrix} \right]
\end{equation}
Here the $j^{th}$ column of \({{\tilde{\bf{F}}}_{i}}\) is given as \(\left\{ \begin{matrix}
   {{{\overset{\lower0.5em\hbox{$\smash{\scriptscriptstyle\frown}$}}{f}}}_{i}}-\tilde{f}_{i}^{\left(j\right)}\\\mathbf{\tilde{x}}_{i}^{{{\mathbf{\sigma}}_{1}}\left(j\right)}-\mathbf{\tilde{x}}_{i}^{\left(j\right)}\\\end{matrix}\right\}\). Let {\it{a}} be a positive real number and define the nearest integer smaller than {\it{a}} by \(\left\lfloor a \right\rfloor \) . Then the integer valued perturbation parameter for the local extremization cum coalescence step may be identified as \(l=\left\lfloor ||{{[{{\mathbf{\gamma }}_{i}}\mathbf{\gamma }_{i}^{T}]}^{-1}}|| \right\rfloor\).
\par A real strength of the coalescence step is in the non-unique choice of the innovation vector - a feature that enables one to design powerful global search schemes. Indeed, one may also borrow some of the basic concepts from different existing global optimization schemes and adapt them within the present coalescence framework. For example, the personal and global information used in the PSO may be incorporated here by constructing the innovation as given below:
\begin{equation} \label{PSO_innov}
\mathbf{\tilde{I}}_{i}^{\left( j \right)}=\left\{ \begin{matrix}
   {}^{p}\mathbf{\tilde{x}}_{i}^{\left( j \right)}-\mathbf{\tilde{x}}_{i}^{\left( j \right)}  \\
   {}^{g}{{{\mathbf{\tilde{x}}}}_{i}}-\mathbf{\tilde{x}}_{i}^{\left( j \right)}  \\
\end{matrix} \right\}
\end{equation}
Here \({}^{p}\mathbf{\tilde{x}}_{i}^{\left( j \right)}\) is the personal best location corresponding to the $j^{th}$ particle along its evolution history till ${\tau _i}$. ${}^g{{\bf{\tilde x}}_i}$ is the available best location among all the particles in the population (ensemble) till ${\tau _i}$, i.e. during the same evolution history. In contrast to the PSO, the gain-type coefficient matrix here is founded on a sound probabilistic basis and hence its matrix structure enables iteration-dependent differential weighting of the scalar components of the innovation vector, thereby yielding faster convergence without losing the exploratory efficacy of the global search. Indeed, it is well recognized that the use of three parameters (e.g. the cognitive and social learning factors and the inertia weight) in the original PSO may lead to solutions that sensitively depend upon the choice of these parameters. Even though it is possible to incorporate within the current setup the basic ideas behind some of the augmented PSO schemes, which attempt at removing some of the shortfalls of the original PSO \cite{22chaotic_prameter_PSO}, such a detailed exercise is kept outside the ambit of the current work. While we do not provide a proof here, the convergence and uniqueness of the iterative increment through this step, for a non-decreasing sequence of $l$  converging to a limit point $l^*$ which may be large, could be shown based on the seminal work of Stroock and Varadhan \cite{16stroock_varadhan} on martingale problems.
\par {\bf{Operator T2: Scrambling and relaxation}}
\par Similar to {\bf{T1}}, the operator {\bf{T2}} also corresponds to random perturbations of the particles in the population. Identifying the gain-like coefficient matrix in the update equation (\ref{update_3}) as a derivative-free stochastic counterpart to the Frechet derivative, the term ${\bf{\tilde C}}_i^{\left( j \right)}$ may be considered the equivalent of a directional derivative term responsible for updating the \(j^{th}\) particle. Consequently, around any extremum, the ${L^2}(P)$ norm $||{\bf{\tilde C}}_i^{\left( j \right)}||$ is likely to be small. This has the effect of rendering further updates of the \(j^{th}\) particle small, leading to a possible stalling of the optimization scheme. In order to move out of these local traps, a possible way is to swap the gain-weighted directional information for the \(j^{th}\) particle with that of another randomly selected one. This form of random perturbation at $\tau_i$  may be implemented by replacing the update equation, e.g. Eqn. (\ref{update4}), by any one of the following two perturbed equations:

\begin{equation} \label{scrambling1}
\mathbf{x}_{i}^{(j)}=\mathbf{\tilde{x}}_{i}^{(j)}+\mathbf{\tilde{D}}_{i}^{{{\mathbf{\sigma }}_{2}}\left( j \right)}
\end{equation}
or
\begin{equation} \label{scrambling2}
\mathbf{x}_{i}^{(j)}=\mathbf{\tilde{x}}_{i}^{{{\mathbf{\sigma }}_{2}}(j)}+\mathbf{\tilde{D}}_{i}^{\left( j \right)}
\end{equation}
where \(\mathbf{\tilde{D}}_{i}^{\left( j \right)}\) is update vector originally computed for the \(j^{th}\) particle via Eqn. (\ref{update4}) and ${{\bf{\sigma }}_2}$ a random permutation on the integer set \(\{1,N\}\). Out of the two perturbed equations as above to implement scrambling, Eqn. (\ref{scrambling2}) is presently adopted. Formally, this perturbation may be described by a probability kernel ${p_l}$ on $[1,N] \times [1,N]$ such that:$\sum\limits_{i \in [1,N]} {{p_l}\left( {i,j} \right)}  = 1\,\,\forall j \in [1,N]$. Clearly, as $l \to \infty $ (or as $\tau  \to \infty$ , i.e. as convergence to the global optimum takes place), the matrix ${p_l}\left( {i,j} \right)$  should ideally approach the identity matrix, i.e. ${p_l}\left( {i,j} \right) \to {\delta _{ij}}$, where ${\delta _{ij}}$  is the Kronecker delta.
\par By borrowing a basic idea from the DE and probably at the cost of a somewhat slower convergence for a class of problems, a more effective modification of the above scrambling strategy may be contemplated with a view to ensuring that the particles explore even more in the search space. Noting that every particle is an {\it{n}}-dimensional vector, one may execute the scrambling operation separately ({\it{n}} times) for individual scalar components of the particles instead of swapping the update terms for the particles as a whole. This may further be followed up by a so called relaxation strategy, wherein the resulting modification is accepted with some probability $\gamma  < 1$. This improvement, enabling element crossovers across different particles, allows the particles to assume larger variations and prompt them to explore more. The resulting update equation is then given by:
\begin{equation} \label{relax}
x_i^{m,(j)} = \tilde x_i^{m,{{\bf{\sigma }}_2}(j)} + \tilde D_i^{m,\left( j \right)},   m \in [1,n]
\end{equation}
where, for instance, $x_i^{m,(j)}$  denotes the updated \(m^{th}\) scalar component of the \(j^{th}\) particle at ${\tau _i}$. Since, upon convergence (and possibly owing to the coalescence step, if applied), all particles crowd around the peak of a unimodal PDF with progressing iterations, directional scrambling with relaxation across such converged particles should not, in any way, affect the numerical accuracy of the estimated global extremum. Hence, in practical implementations of our scheme, ${p_l}\left( {i,j} \right)$ need not strictly approach the identity matrix for large {\it{l}}.
\par {\bf{Operator T3: Selection}}
\par Use of diffusion-based random perturbations during exploration might sometimes result in `bad' candidates. This may necessitate a selection step wherein candidates for the next iteration (say, the \(i^{th}\) iteration) are chosen based on some selection criteria quantified by a selection or fitness function $g({\mathbf{\upsilon }}|{{\bf{x}}_{i - 1}})$, ${\bf{\upsilon }} \in \Omega $. A general construction of the function, which is a Markov transition kernel on $\Omega $  and is conditioned on the ensemble of particles at the last iteration, should satisfy the following properties:
\par a) $g({\bf{\upsilon }} = {\bf{x}}_i^{(j)}|{{\bf{x}}_{i - 1}} = {\bf{x}}_{i - 1}^{(k)}) = 0{\text{  if  }}j \ne k;{\text{ }}\forall j,k \in [1,N]{\text{  }}$
\par b) $g({\bf{\upsilon }} = {\bf{x}}_i^{(j)}|{{\bf{x}}_{i - 1}} = {\bf{x}}_{i - 1}^{(j)},f({\bf{x}}_i^{(j)}) \geqslant f({\bf{x}}_{i - 1}^{(j)})) = \varsigma {\text{ }}\forall j \in [1,N]{\text{  where  }}\varsigma  \in (0,1]$
\par c) $g({\bf{\upsilon }} = {\bf{x}}_{i - 1}^{(j)}|{{\bf{x}}_{i - 1}} = {\bf{x}}_{i - 1}^{(j)},f({\bf{x}}_i^{(j)}) < f({\bf{x}}_{i - 1}^{(j)})) = \varsigma {\text{ }}$
\par The updated \(j^{th}\) particle appearing in the above clauses is computed using Eqn. (\ref{scrambling1}) (or Eqn. (\ref{scrambling2})), which may incorporate different combinations of the three operations, e.g. local extremization, coalescence and scrambling-cum-relaxation. For the operator {\bf{T3}}, the integer-valued inverse-perturbation parameter {\it{l}}  may be identified with $l = \left\lfloor {\frac{1}{{1 - \varsigma }}} \right\rfloor $. In the current numerical implementations of COMBEO as described in Section {\ref{algorithm}},  $\varsigma $=1 is consistently adopted. This corresponds to {\it{l}}  being infinity across all iterations and implies that the selection procedure is deterministic.
\par This work has primarily been aimed at the proposal for a class of new evolutionary optimization schemes and a verification of its performance purely through the numerical route. Accordingly, a detailed convergence analysis of the asymptotic dynamics, based on a combination of the martingale theory of Stroock-Varadhan and the random perturbation theory of Freidlin-Wentzell, will be considered in a separate study.
\section{Algorithm development} \label{algorithm}
The local search and random perturbation approaches, as described in Sections {\ref{martingale_problem}} and {\ref{coalescence_scrambling}}, merely provide a set of general tools whose different combinations could lead to different evolutionary schemes for optimization. For instance, incorporation of the innovation term in Eqn. (\ref{extremal_eq1}) yields a greedy algorithm that, despite possibly faster convergence, is likely to be a poor performer in the global search. On the other hand, to facilitate a more exhaustive global search at the cost of a substantively slower convergence, only a PSO-type innovation as in Eqn. (\ref{PSO_innov}) could be adopted. For better clarity and a more objective assessment of the strengths and demerits of different search tools, three pseudo-codes are presented in this section. A more ambitious alternative would have been to combine all the presented ideas in a single pseudo-code that could automatically offer a right mix of convergence speed with exploratory efficiency depending on the nature of the problem at hand. This exercise however needs a non-trivial extension of the current work and hence constitutes an interesting future problem at this stage. {\bf{\it{pseudo-code 1}}} uses the innovation term of Eqn. (\ref{extremal_eq1}) for a greedy local search whilst employing scrambling of particles as a whole (not element wise) for the global search. Importantly, it may be noted that while the prediction step of Eqn. (\ref{process_eq}) appears to be helpful in exploration, this step is not practically useful with our schemes, especially as it does not exploit any directional information and as more efficient tools for explorations have already been laid out. Hence, given a random initial scatter to the particles provided at the beginning of the iterations, the prediction step has been completely eliminated from all subsequent iterations. This would typically mean that the number of evaluations of the objective functional would be reduced by half. Note that, as there is no prediction step, the over-tildes in the notations of the variables are henceforth removed.

\par {{\bf{COMBEO} }\emph{pseudo-code 1:}}
\par 1. Discretize the  $\tau $-axis, say $\left[ {{\tau _{\min }},{\tau _{\max }}} \right]$, using a partition $\left\{ {{\tau _0} = {\tau _{\min }},{\tau _1},...,{\tau _M} = {\tau _{\max }}} \right\}$ such that ${\tau _0} < ... < {\tau _M}$  and  ${\tau _i} - {\tau _{i - 1}} = \Delta {\tau _i}{\text{ }}( = \Delta \tau : = \frac{1}
{M}$ if a uniform step-size is chosen for $i = 0,...,M - 1$ ). Assign ${\tau _0} = 1$ and adopt $\Delta \tau $ small ($ \sim {10^{ - 7}}$). Choose an ensemble size {\it{N}}.
\par 2.	From the domain of definition randomly generate, following a uniform distribution, the ensemble of initial particles (the initial population) $\left\{ {{\bf{x}}_0^{(j)}} \right\}_{j = 1}^N$ for the solution vector. For each discrete ${\tau _i},{\text{ }}i = 1,...,M - 1$, execute the following steps.
\par 3.	{\it{Additive update }}
\par Choose  $\alpha  \in (0,1)$; a typical prescription could be $\alpha  \approx 0.8$, even though the method generally performs well for other values in the interval indicated. Update each particle as:
$${\bf{x}}_i^{(j)} = {\bf{x}}_{i - 1}^{{{\bf{\sigma }}_2}\left( j \right)} + {\bf{D}}_i^{\left( j \right)},\,\,\,j = 1,...,N$$
where ${\bf{D}}_i^{\left( j \right)}$ is the  \(i^{th}\) correction vector. ${{\bf{\sigma }}_2}\left( j \right)$ is the \(j^{th}\) random permutation based on a uniformly distributed probability measure over the integer set$\left\{ {1,...,N} \right\}$. For convenience, the expression for ${\bf{\tilde D}}_i^{\left( j \right)}$ is reproduced below:
$${\bf{D}}_i^{\left( j \right)}: = {\beta _i}{{\bf{G}}_i}{\bf{I}}_i^{\left( j \right)}\,\,\,,j = 1,...,N$$
\[\mathbf{I}_{i}^{\left( j \right)}:=\left\{ \begin{matrix}
   {{{\overset{\lower0.5em\hbox{$\smash{\scriptscriptstyle\frown}$}}{f}}}_{i-1}}-f_{i-1}^{\left(j\right)}\\\mathbf{x}_{i-1}^{{{\mathbf{\sigma}}_{1}}\left(j\right)}-\mathbf{x}_{i-1}^{\left(j\right)}\\\end{matrix}\right\}\]
${\beta _i} = {\hat \beta _i}{\zeta _i}$ (${\hat \beta _i}$ is a scalar constant and ${\zeta _i}$ a uniform random number between 0 and 1. ${{\bf{\sigma }}_1}\left( j \right)$ is defined as the \(j^{th}\) candidate from another (independent) random permutation of the integer set $\left\{ {1,...,N} \right\}$. Finally,

\[\begin{split}{{\bf{G}}_i}: = \frac{1}{N}\left\{\begin{array}{c}{\left( {{{\bf{X}}_{i - 1}} - {{\hat {\bf{X}}}_{i - 1}}} \right)\left( {{{\bf{F}}_{i - 1}}^T{\tau _i} - {{\hat {\bf{F}}}_{i - 1}}^T{\tau _{i - 1}} - \Delta {{\hat {\bf{F}}}_{i - 1}}^T{\tau _i}} \right)}\\{ + \left( {{{\hat {\bf{X}}}_{i - 1}}{\tau _i} - {{\hat {\bf{X}}}_{i - 1}}{\tau _{i - 1}}} \right){{\left( {{{\bf{F}}_{i - 1}} - {{\hat {\bf{F}}}_{i - 1}}} \right)}^T}}\end{array}\right\}\\\left\{\begin{array}{c}{\alpha \frac{1}
{{N - 1}}\left( {\left( {{{\overset{\lower0.5em\hbox{$\smash{\scriptscriptstyle\frown}$}}{{{\bf{F}}}}_{i - 1} }} - {{\bf{F}}_{i - 1}}} \right) - \left( {{{\overset{\lower0.5em\hbox{$\smash{\scriptscriptstyle\frown}$}}{{{\bf{F}}}}_{i - 1} }} - {{\hat {\bf{F}}}_{i - 1}}} \right)} \right)}\\{{{\left( {\left( {{{\overset{\lower0.5em\hbox{$\smash{\scriptscriptstyle\frown}$}}{{{\bf{F}}}}_{i - 1} }} - {{\bf{F}}_{i - 1}}} \right) - \left( {{{\overset{\lower0.5em\hbox{$\smash{\scriptscriptstyle\frown}$}}{{{\bf{F}}}}_{i - 1} }} - {{\hat {\bf{F}}}_{i - 1}}} \right)} \right)}^T} + \left( {1 - \alpha } \right){{\bf{\gamma }}_i}{\bf{\gamma }}_i^T}\end{array}\right\}^{-1}\end{split}\]
\par 4.	If $f({\bf{x}}_i^{(j)}) \geqslant f\left( {{\bf{x}}_{i - 1}^{\left( j \right)}} \right)\,,\,j = 1,...,N,$   then retain ${\bf{x}}_i^{(j)}$ as the updated particle;
\par else set ${\bf{x}}_i^{(j)} = {\bf{x}}_{i - 1}^{(j)}$.
\par 5.	 If $i < M$, go to Step 3 with $i = i + 1$,
else terminate the algorithm and report \(\frac{1}{N}\sum\limits_{j = 1}^N {{\bf{x}}_i^{(j)}}\) as an estimate for the global optimum.
\par In the next pseudo-code, the DE-like scalar element-wise scrambling scheme for particles is implemented and the resulting modification accepted with some relaxation criteria. The greedy local search based on the innovation term ${\overset{\lower0.5em\hbox{$\smash{\scriptscriptstyle\frown}$}}{{f}}_{i - 1} } - f_{i - 1}^{\left( j \right)}$ is not included in this pseudo-code as it might negatively bear on the global exploration. It may be possible to modify the pseudo-code to augment the innovation vector with additional innovation terms like ${\overset{\lower0.5em\hbox{$\smash{\scriptscriptstyle\frown}$}}{{f}}_{i - 1} } - f_{i - 1}^{\left( j \right)}$  after properly weighing the different terms. However, such a variation is presently not attempted.

\par {{\bf{COMBEO} }\emph{pseudo-code 2:}}

\par 1.	Replicate steps 1 and 2 from {\emph{pseudo-code 1}}.
\par 2.	{\emph{Additive update}}
\par Choose $\alpha  \in (0,1)$  as in {\emph{pseudo-code 1}} and update each particle as described below.
Draw a random number {\emph{r}} uniformly from the set $\left\{ {1,...,n} \right\}$. As in {\emph{pseudo-code 1}}, define ${{\bf{\sigma }}_2}\left( j \right)$  as the \(\it{j^{th}}\) candidate from a uniformly random permutation over the integer set$\left\{ {1,...,N} \right\}.$
\par Initialize: ${\bf{x}}_i^{(j)} = {\bf{x}}_{i - 1}^{{{\bf{\sigma }}_2}\left( j \right)}\,\,\,,\,\,\,\,\,\,j = 1,...,N$ .
\par If  ${\zeta _i} < {c_i}$ ($0 < {c_i} \leqslant 1$ is a scalar chosen by the end-user and ${\zeta _i}$ an independent uniform random variable in $[0,1]$), then
$$x_i^{r,(j)} = x_{i - 1}^{r,{{\bf{\sigma }}_2}\left( j \right)} + D_i^{r,\left( j \right)},\,\,\,j = 1,...,N$$
\par with ${\bf{D}}_i^{\left( j \right)}$ as in {\emph{pseudo-code 1}} and  ${\bf{I}}_i^{\left( j \right)}: = \left\{ {{\bf{x}}_{i - 1}^{{{\bf{\sigma }}_1}\left( j \right)} - {\bf{x}}_{i - 1}^{\left( j \right)}} \right\}$, where ${{\bf{\sigma }}_1}\left( j \right)$  is a uniformly random permutation on the index set $\{ 1,N\} \backslash \{ j\} $;
\par else
$$x_i^{m,(j)} = x_{i - 1}^{m,{{\bf{\sigma }}_2}\left( j \right)}$$
\par Set ${\emph{r}} \to {\emph{r}} + 1$. Run the loop on {\emph{r}} till it reaches {\emph{n}}.
\par 3.	Replicate steps 4 and 5 from {\emph{pseudo-code 1}}.
\par By using the available personal and global best information as in the PSO, one may propose yet another global optimization pseudo-code, as suggested earlier while discussing the coalescence step.  Thus the search here is based purely on a martingale problem without requiring a random perturbation, especially the scrambling step. This is an interesting proposal in the sense that the search for the global optimum is attempted by just constructing an innovation vector and driving the same to a zero-mean martingale. In the previous two algorithms ({\emph{pseudo-code 1}} and {\emph{pseudo-code 2}}), the update terms were not constructed using past information beyond one step. In the {\emph{pseudo-code 3}} presented below, the update term is however based on the evolution history till the last iteration in order to explore the search space more exhaustively. Without a loss of generality, the specific algorithm given below is for a global maximization problem.
\par {{\bf{COMBEO} }\emph{pseudo-code 3:}}
\par 1.	Same as in {\emph{pseudo-code 1}}.
\par 2.	Generate the initial population $\left\{ {{\bf{x}}_0^{(j)}} \right\}_{j = 1}^N$.
\par Initialize: ${\vartheta _{\max }} = 1$ and ${\vartheta _{\min }} = 0.1$ .
\par  Initialize the local and global best particles ${}^p{\bf{x}}_0^{\left( j \right)} = {\bf{x}}_0^{\left( j \right)},\,\,\,j = 1,...,N$ and ${}^g{{\bf{x}}_0} = \arg \max \left\{ {f({\bf{x}}_0^{(1)}),...,f({\bf{x}}_0^{(N)})} \right\}$. Initialize the correction term ${\bf{D}}_0^{\left( j \right)} = {\bf{0}},\,\,j = 1,...,N.$  For each discrete ${\tau _i},{\text{ }}i = 1,...,M - 1$, execute the following steps.

\par 3.	{\emph{Additive update}}
\par Choose $\alpha  \in (0,1)$
\par Compute ${\vartheta _i} = {\vartheta _{\max }} - \left( {{\vartheta _{\max }} - {\vartheta _{\min }}} \right)\frac{i}
{M}$
\par  Update each particle as:
$${\bf{x}}_i^{(j)} = {\bf{x}}_{i - 1}^{\left( j \right)} + {\bf{D}}_i^{\left( j \right)},\,\,\,j = 1,...,N$$
\par  ${\bf{D}}_i^{\left( j \right)}$ is the  ${\it{j^{th}}}$ update vector. The expression for ${\bf{D}}_i^{\left( j \right)}$ is also reproduced below:
$${\bf{D}}_i^{\left( j \right)}: = {\vartheta _i}{\bf{D}}_{i - 1}^{\left( j \right)} + {\beta _i}{{\bf{G}}_i}{\bf{I}}_i^{\left( j \right)},{\text{  }}j = 1,...,N$$,
\[\mathbf{\tilde{I}}_{i}^{\left( j \right)}=\left\{ \begin{matrix}
   \mathbf{\tilde{x}}_{i}^{p\left( j \right)}-\mathbf{\tilde{x}}_{i}^{\left( j \right)}  \\
   \mathbf{\tilde{x}}_{i}^{g}-\mathbf{\tilde{x}}_{i}^{\left( j \right)}  \\
\end{matrix} \right\}\]
\par  ${\beta _i} = {\hat \beta _i}{\zeta _i}$ (${\hat \beta _i}$
is a scalar constant and ${\zeta _i}$ a uniform random number in $[0,1]$).
\par 4.	Update the personal and global information as:
\par if $f({\bf{x}}_i^{(j)}) \geqslant f\left( {{\bf{x}}_{i - 1}^{\left( j \right)}} \right)\,,\,j = 1,...,N,$   then retain ${}^p{\bf{x}}_i^{(j)} = {\bf{x}}_i^{(j)}$ as the updated particle;
\par if $\max \left\{ {f({\bf{x}}_i^{(1)}),...,f({\bf{x}}_i^{(N)})} \right\} \geqslant f\left( {{}^g{{\bf{x}}_{i - 1}}} \right)\,$   then ${}^g{{\bf{x}}_i} = \arg \,\max \left\{ {f({\bf{x}}_i^{(1)}),...,f({\bf{x}}_i^{(N)})} \right\}$ ;
\par else ${}^g{{\bf{x}}_i} = {}^g{{\bf{x}}_{i - 1}}$ .

\par Before concluding this section, a word of caution regarding the claims made on a relatively superior performance of the present approach vis-a-vis a few others, e.g. the DE and the PSO, should be in order. Optimization schemes like the DE or the PSO have been extensively improved over the last 15 years or so and it would be wrong to qualify the current method (as captured through the pseudo-codes above) as being the most competent. Nevertheless, the inherent flexibility of our proposal should permit the incorporation of the basic concepts behind some of the recent improvements in the DE or the PSO \cite{survey_DE_imprv, DE_survey, DE_3, self_learning_PSO}, thereby rendering the latter even more effective. Such details however need to be worked out and do not form part of the present study.

\section{Numerical illustrations}
\subsection{Benchmark problems}
\par While the efficiency of a global optimization strategy is largely dependent on its ability to search all the promising search regions, the complexity level of a given optimization problem could exponentially increase with increasing number of design variables. As an example, even though Rosenbrock's function is unimodal in 2 dimensions ($n = 2$ ), it is highly multimodal in still higher dimensional search spaces. Hence, a global optimization scheme, which works successfully for lower dimensional problems, may very well fail as the dimension of the problem increases. Moreover, the level of difficulty in finding the global optimum may also be dependent on the specific characteristics of the objective function, which goes to explain the varied degree of difficulty in solving different optimization problems of the same dimension. If the problem is separable or partially separable, i.e. if the objective function can be additively split in terms of component functions each of which is expressible in terms of just one element or a small subset of elements of the design variable vector {\bf{x}}, the original problem may actually be decoupled into a set of sub-problems. Each sub-problem, involving one scalar variable or a small set of scalar variables, may be solved separately, e.g. by a simple line search for one scalar variable. Depending on the degree of complexity, optimization problems could be categorized as separable, $m$-non-separable and non-separable {\cite{23benchmark_functions}}. In between the two extreme cases i.e. separable and non-separable, are the $m$-non-separable functionals ($m$ being the maximum number of scalar design variables appearing in the descriptions of the component functions) that correspond to partially separable problems. The nomenclature `{\emph{separability}}' bears a similar meaning as `{\emph{epistasis}}' in biology. In this work, in testing the effectiveness of the proposed schemes to some extent, the following problems have been considered.
\par I. Separable Functions:
\par (a) {\emph{F}}1: Shifted Elliptic Function
\par (b) {\emph{F}}2: Shifted Rastrigin's Function
\par (c) {\emph{F}}3: Shifted Ackley's Function

\par II. Single-group {\emph{m}}-non-separable Functions
\par (d) {\emph{F}}4: Single-group Shifted and {\emph{m}}-rotated Elliptic Function
\par (e) {\emph{F}}5: Single-group Shifted and {\emph{m}}-rotated Rastrigin's Function
\par (f) {\emph{F}}6: Single-group Shifted {\emph{m}}-dimensional Schwefel's Problem 1.2
\par (g) {\emph{F}}7: Single-group Shifted {\emph{m}}-dimensional Rosenbrock's Function

\par III. $\frac{n}{{2m}}$-group {\emph{m}}-non-separable Functions
\par (h) {\emph{F}}8:  $\frac{n}{{2m}}$-group Shifted and {\emph{m}}-rotated Rastrigin's Function
\par (i) {\emph{F}}9:   $\frac{n}{{2m}}$-group Shifted and {\emph{m}}-rotated Schwefel's Problem 1.2

\par IV.  $\frac{n}{{m}}$-group {\emph{m}}-non-separable Functions
\par (j) {\emph{F}}10:   $\frac{n}{{m}}$-group Shifted and {\emph{m}}-rotated Schwefel's Problem 1.2

\par V. Non-separable Functions
\par (k) {\emph{F}}11: Shifted Schwefel's Problem 1.2
\par The above functions are constructed using the following basic functions.
\par B1. The Sphere Function: ${F_{sphere}}\left( {\bf{x}} \right) = \sum\limits_{j = 1}^n {{{\left( {{x^j}} \right)}^2}} $
\par B2. The Elliptic Function: ${F_{elliptic}}\left( {\bf{x}} \right) = \sum\limits_{j = 1}^n {\left[ {{{\left( {{{10}^6}} \right)}^{\frac{{j - 1}}
{{n - 1}}}}{{\left( {{x^j}} \right)}^2}} \right]} $
\par B3. The Rotated Elliptic Function: ${F_{rot\_elliptic}}\left( {\bf{x}} \right) = {F_{elliptic}}\left( {\bf{z}} \right)$ , ${\bf{z}} = {\bf{x}} * {\bf{M}}$ ({\bf{M}} being an orthogonal matrix)
\par B4. Schwefel's Problem 1.2: ${{F}_{schwefel}}\left( \mathbf{x} \right)=\sum\limits_{k=1}^{n}{{{\left( \sum\limits_{j=1}^{k}{{{x}^{j}}} \right)}^{2}}}$
\par B5. Rosenbrock's Function: ${{F}_{rosenbrock}}\left( \mathbf{x} \right)=\sum\limits_{j=1}^{n-1}{\left[ 100{{\left( {{\left( {{x}^{j}} \right)}^{2}}-{{x}^{j+1}} \right)}^{2}}+{{\left( {{x}^{j}}-1 \right)}^{2}} \right]}$

\par B6. Rastrigin's Function: ${{F}_{rastrigin}}\left( \mathbf{x} \right)=\sum\limits_{j=1}^{n}{\left[ {{\left( {{x}^{j}} \right)}^{2}}-10\cos \left( 2\pi {{x}^{j}} \right)+10 \right]}$

\par B7. Rotated Rastrigin's Function: ${{F}_{rot\_rastrigin}}\left( \mathbf{x} \right)={{F}_{rastrigin}}\left( \mathbf{z} \right)$, $\mathbf{z}=\mathbf{x}*\mathbf{M}$ ({\bf{M}} being a orthogonal matrix)

\par B8. Ackley's Function :  		${{F}_{ackley}}\left( \mathbf{x} \right)=-20exp\left( -0.2\frac{1}{n}\sqrt{\sum\limits_{j=1}^{n}{{{\left( {{x}^{j}} \right)}^{2}}}} \right)-\exp \left( \frac{1}{n}\sum\limits_{j=1}^{n}{\cos \left( 2\pi {{x}^{j}} \right)} \right)+20+\exp \left( 1 \right)$

\par B9. Rotated Ackley's Function: ${{F}_{rot\_ackley}}\left( \mathbf{x} \right)={{F}_{ackley}}\left( \mathbf{z} \right)$, $\mathbf{z}=\mathbf{x}*\mathbf{M}$

\par Explicit expressions for the functions $F1$-$F11$ are given below. Let $\mathbf{z}=\mathbf{x}-\mathbf{o}$. $\mathbf{o}\in {{\mathbb{R}}^{n}}$ denote a shifted global optimum  and ${{\sigma }_{n}}$ the random permutation of the vector $\left\{ 1,...,n \right\}$.

\par (a) $F1\left( \mathbf{x} \right)={{F}_{elliptic}}\left( \mathbf{z} \right)$
\par (b) $F2\left( \mathbf{x} \right)={{F}_{rastrigin}}\left( \mathbf{z} \right)$
\par (c) $F3\left( \mathbf{x} \right)={{F}_{ackley}}\left( \mathbf{z} \right)$
\par (d) $F4\left( \mathbf{x} \right)={{F}_{rot\_elliptic}}\left( \mathbf{z}\left( {{\sigma }_{n}}\left( 1:m \right) \right) \right)\times {{10}^{6}}+{{F}_{elliptic}}\left( \mathbf{z}\left( {{\sigma }_{n}}{{\left( m+1:n \right)}} \right) \right)$
\par (e) $F5\left( \mathbf{x} \right)={{F}_{rot\_rastrigin}}\left( \mathbf{z}\left( {{\sigma }_{n}}\left( 1:m \right) \right) \right)\times {{10}^{6}}+{{F}_{rastrigin}}\left( \mathbf{z}\left( {{\sigma }_{n}}{{\left( m+1:n \right)}} \right) \right)$
 \par (f) $F6\left( \mathbf{x} \right)={{F}_{schwefel}}\left( \mathbf{z}\left( {{\sigma }_{n}}\left( 1:m \right) \right) \right)\times {{10}^{6}}+{{F}_{sphere}}\left( \mathbf{z}\left( {{\sigma }_{n}}{{\left( m+1:n \right)}} \right) \right)$
\par (g) $F7\left( \mathbf{x} \right)={{F}_{rosenbrock}}\left( \mathbf{z}\left( {{\sigma }_{n}}\left( 1:m \right) \right) \right)\times {{10}^{6}}+{{F}_{sphere}}\left( \mathbf{z}\left( {{\sigma }_{n}}{{\left( m+1:n \right)}} \right) \right)$
\par (h) $F8\left( \mathbf{x} \right)=$\par $\sum\limits_{k=1}^{\frac{n}{2m}}{ { \begin{bmatrix}{{F}_{rot\_rastrigin}}\left( \mathbf{z}\left( {{\sigma }_{n}}\left( \left( k-1 \right)m+km \right) \right) \right)\times {{10}^{6}}\\+{{F}_{rastrigin}}\left( \mathbf{z}\left( {{\sigma }_{n}}{{\left( \frac{n}{2}+1:n \right)}} \right) \right)\end{bmatrix} } }$
\par (i)$F9\left( \mathbf{x} \right)=$
\par $\sum\limits_{k=1}^{\frac{n}{2m}}{ \begin{bmatrix}{{F}_{schwefel}}\left( \mathbf{z}\left( {{\sigma }_{n}}\left( \left( k-1 \right)m+km \right) \right) \right)\times {{10}^{6}}\\+{{F}_{sphere}}\left( \mathbf{z}\left( {{\sigma }_{n}}{{\left( \frac{n}{2}+1:n \right)}} \right) \right)\end{bmatrix} }$
\par (j) $F10\left( \mathbf{x} \right)=\sum\limits_{k=1}^{\frac{n}{m}}{{{F}_{schwefel}}\left( \mathbf{z}\left( {{\sigma }_{n}}\left( \left( k-1 \right)m+km \right) \right) \right)}$
\par (k) $F11\left( \mathbf{x} \right)={{F}_{schwefel}}\left( \mathbf{z} \right)$

\begin{table}[ht]
\caption{}
\begin{tabular}{|c|c|c|c|c|}
\hline
\multicolumn{1}{|c|}{O.F.} & \multicolumn{2}{|c|}{{\emph{pseudo-code 2}}} & \multicolumn{2}{|c|}{DE}\\
\cline{2-3} \cline{4-5} {} & \multicolumn{1}{c|}{N.I.} & \multicolumn{1}{|c|}{M.E.N.} & \multicolumn{1}{|c|}{N.I.} & \multicolumn{1}{|c|}{M.E.N.}\\
\hline
$F1$ & $1.69\times {{10}^{3}}$ & $\varepsilon$ & $2.42\times {{10}^{4}}$ & $\varepsilon $\\
$F2$ & $4.68\times {{10}^{3}}$ & $\varepsilon$ & $2.16\times {{10}^{4}}$ & $\varepsilon $\\
$F3$ & $3.73\times {{10}^{3}}$ & $\varepsilon$ & $it_{max}$ & 20\\
$F4$ & $1.34\times {{10}^{4}}$ & $\varepsilon$ & $it_{max}$ & $1.2\times {{10}^{4}}$\\
$F5$ & $3\times {{10}^{3}}$ & $\varepsilon$ & $2.37\times {{10}^{4}}$ & $\varepsilon$\\
$F6$ &	2.51$\times {{10}^{3}}$ &	$\varepsilon $ &	2.12$\times {{10}^{4}}$ &	$\varepsilon $\\
$F7$ &	1.11$\times {{10}^{4}}$ &	$\varepsilon $ &	2.81$\times {{10}^{4}}$ &	$\varepsilon $\\
$F8$ &	4.15$\times {{10}^{4}}$	&   $\varepsilon $ &	5.84$\times {{10}^{4}}$ &   $\varepsilon $\\
$F9$ &	2.29$\times {{10}^{3}}$	&   $\varepsilon $ &	3.62$\times {{10}^{4}}$	&   $\varepsilon $\\
$F10$&	8.84$\times {{10}^{3}}$&	$\varepsilon $	&$\text{it }\!\!\_\!\!\text{ max}$&	575\\
$F11$&	$\text{it }\!\!\_\!\!\text{ max}$ &	7.7$\times {{10}^{-4}}$	&$\text{it }\!\!\_\!\!\text{ max}$ &	5.78 $\times {{10}^{3}}$\\
\hline
\end{tabular}
\end{table}
In table 1 and 2, O.F., N.I. and M.E.N. are used to denote objective functional, number of iterations and mean error norm respectively.
\begin{table}[ht]
\caption{}
\begin{tabular}{|c|c|c|c|c|}
\hline
\multicolumn{1}{|c|}{O.F.} & \multicolumn{2}{|c|}{{\emph{pseudo-code 3 (pseudo-code  1)}}} & \multicolumn{2}{|c|}{PSO}\\
\cline{2-3} \cline{4-5} {} & \multicolumn{1}{c|}{N.I.} & \multicolumn{1}{|c|}{M.E.N.} & \multicolumn{1}{|c|}{N.I.} & \multicolumn{1}{|c|}{M.E.N.}\\
\hline
B1 & 146(37) & $\varepsilon$ & 330 & $\varepsilon $\\
B2 & 172 (55) & $\varepsilon$ & 353 & $\varepsilon $\\
B3 & 212 (64) & $\varepsilon$ & 630 & $\varepsilon$\\
B4 & 190 (49) & $\varepsilon$ & 456 & $\varepsilon$\\
B5 & 154 (35) & $\varepsilon$ & 280 & $\varepsilon$\\
B6 &	155 (42) &	$\varepsilon $ & 353 &	$\varepsilon $\\
B7 &	172 (46) &	$\varepsilon $ &	330 &	$\varepsilon $\\
B8 &	195 (61)	&   $\varepsilon $ &	485 &   $\varepsilon $\\
B9 &	201 (59)	&   $\varepsilon $ &	440	&   $\varepsilon $\\
\hline
\end{tabular}
\end{table}
In the numerical work reported here, we have consistently used $n=40$. In solving optimization problems involving functions $F$1 - $F$11, the performance of \emph{pseudo-code 2}, as compared with the parent DE, is given in Table I. The population set in both the cases consists of 2000 particles. The scalar parameter $c_i$, which is similar to the acceptance criteria CR in DE \cite{3storn_DE} is taken as 0.1. In implementing DE, CR is also taken as 0.1. For both the schemes, the program is terminated if the number of iteration exceeds the maximum allowed threshold, which is presently given by $\text{it }\!\!\_\!\!\text{ max}=1\times {{10}^{5}}$. In all the tables, $\varepsilon$ denotes an error norm $\le {{10}^{-5}}$. By way of investigating the consistent reproducibility of the results reported herein, the proposed optimization scheme ({\emph{pseudo-code 2}}) and the DE are run 5 times for $F$3. It is observed that in all the cases {\emph{pseudo-code 2}} provides the optimum in roughly the same number of iterations (same range) mentioned in Table I, whereas the DE fails to converge in all the runs. A typical evolution of the mean of the cost functional $F$3 via {\emph{pseudo-code 2}} is presented in figure (\ref{cost_functional}). In the literature, the above benchmark problems are typically considered to be `difficult' to solve by most of the existing evolutionary optimization schemes \cite{23benchmark_functions}.

\begin{figure}[!h]
    \begin{center}
            \label{1}
            \includegraphics[width=1\textwidth]{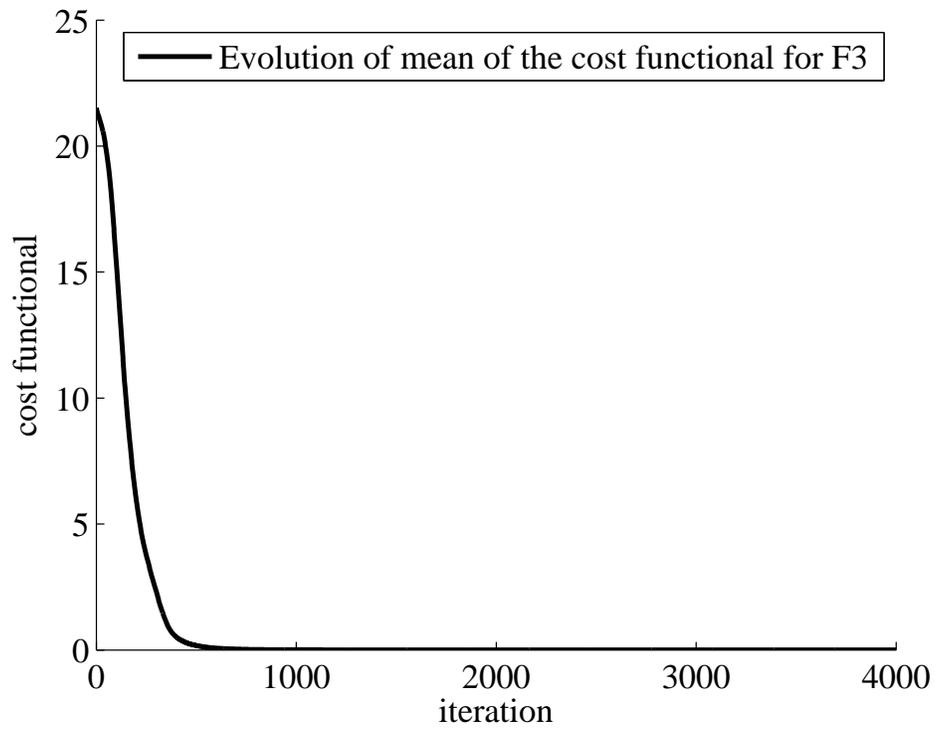}
    \end{center}
    \caption{%
        Evolution of mean of the cost functional for $F$3 by {\emph{pseudo-code 2}}
     }%
     \label{cost_functional}
\end{figure}

Indeed, for these problems, the PSO, {\emph{ pseudo-code 1}} and {\emph{ pseudo-code 3}} could not converge to the global optimum within the presently allowed number of iterations. However, in order to demonstrate that {\emph{ pseudo-code 1 and 3}} could indeed be very accurate with faster convergence for many global optimization problems, these codes are tested with benchmark global optimization problems B1-B9. While these problems are also challenging in their own right and often used in the literature to assess the performance of optimization algorithms, they do pose comparatively lesser degree of complexity in reaching the global optimum vis-a-vis $F$1-$F$11. The results, reported in Table II, are also compared against the PSO under a similar environment. In all the cases, the population set is taken to consist of 50 particles. As anticipated, {\emph{ pseudo-code 3}} and especially {\emph{ pseudo-code 1}} show conspicuously faster convergence than the PSO whilst maintaining the same or higher level of accuracy. Nevertheless, there is clearly scope for improving these two pseudo-codes so that they can be rendered as competitive as {\emph{ pseudo-code 2}}. In exploring modification over the present algorithms so as to render them more robust and applicable to higher dimensional optimization problems, one possible way, for instance, would be {\emph{intelligent}} augmentation of multiple adaptive strategies \cite{25adaptivePSO}.  Moreover, since there are no specific formulas for selecting the parameter values in the present algorithms, parameter control mechanisms within the martingale problem set-up may also have to be suitably devised \cite{25adaptivePSO}.

\subsection{An application to quantitative estimation of density variation in high-speed flows through inversion of light travel-time data}

Having assessed the proposed global optimization framework through a variety of benchmark problems, we now consider a considerably larger dimensional optimization problem that involves the quantitative estimation of density variation in a high-speed flow obstructed by a blunt-nose aerodynamic vehicle. This example is taken up to demonstrate that, while {\it pseudo-code 2} often provides an effective search tool for the global extremum of complex objective functionals, a greedier search scheme (as in {\it pseudo-code 1} or its possible variants) could nevertheless be more appropriate for many practically useful, yet large dimensional, problems where a scheme like {\it pseudo-code 2} (or most available global schemes such as the DE or the PSO) might prove computationally prohibitive.

\begin{figure}[!h]
    \begin{center}
            \label{01}
           \includegraphics[width=0.5\textwidth]{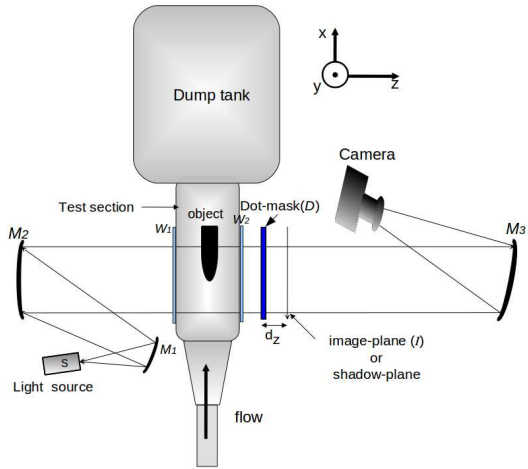}
    \end{center}
    \caption{%
        A 3D view of the reconstruction via the proposed framework
     }%
     \label{exp_setup}
\end{figure}
  The experimental set-up is in figure (\ref{exp_setup}); see \cite{opt} for more details. Here, the region of interest  (ROI), the flow around the object (in this case, a blunt-nosed missile model), is illuminated by a plane wave. The distorted plane wave trans-illuminates a random dot pattern (RDP). The RDP in effect serves the purpose of the expensive lenslet array Shack-Hartmann sensor. The geometric shadow cast at an adjacent plane is imaged by a high-speed camera. The shadow, in comparison to the original RDP, is space-shifted. Local mean shifts $\boldsymbol{d}=(d_x,d_y)$ and a related quantity, the slopes of the wavefront $\boldsymbol{\theta =}\left( {\tfrac{d_x}{d_z},\tfrac{d_y}{d_z}} \right)$, are estimated by cross-correlating local sub-region of the distorted RDP with the corresponding original and finding the shift of the cross-correlation peak. The estimated local slopes are integrated to find the smooth wavefront $\Phi$ \cite{opt2}. The recovered unwrapped $\Phi$, which carries the distortion owing to its passage through the ROI, is shown in figure (\ref{recon1}). \\
  The distorted wavefront $\Phi^i$ (or equivalently the phase delay) evaluated at the $i^{th}$ detector, is related to the refractive index distribution $\varrho(\mathbf{r})$ in the ROI, $\mathbf{r}=(x,y,z)$ being the position vector, through the equation:
  $${{\Phi^i }}=\frac{2\pi }{\lambda }\int\limits_{0}^{L^i}{\varrho(\mathbf{r})ds}$$
  Here $L^i$ is the length of the $i^{th}$ ray and $ds$ is an element from the Fermat's path of the light-ray through $\varrho(\mathbf{r})$, which obeys the Eikonel equation:
  $$\frac{d}{ds}\left( \varrho\frac{d\mathbf{r}}{ds} \right)=\nabla \varrho$$

  Therefore the equation connecting $\Phi^i$ to $\varrho(\mathbf{r})$ is nonlinear. The collection of $\Phi^i$ at all the detectors is denoted using the vector $\boldsymbol{\Phi}:=\left\{\Phi^i\right\}$. The problem of recovery of $\varrho(\mathbf{r})$ from the experimentally measured data, denoted by  $\boldsymbol{\Phi}_{\text{exp}}$, obtained from $\Phi$ corresponding to the locations of detectors, is posed as an optimization problem which is solved within the proposed framework. The corresponding objective functional may be written as:

$$f_{\text{exp}}:=\sqrt{(\boldsymbol{\Phi}_{\text{exp}}-\boldsymbol{\Phi})^T(\boldsymbol{\Phi}_{\text{exp}}-\boldsymbol{\Phi})}$$
   However, given the large system dimension ($n=101^2$) and with the algorithm requiring repetitive numerical inversions for $\Phi$, most available global optimization schemes (e.g. the DE or the PSO) are rendered impracticable using the commonly available computing systems. An acceleration of the computing speed is however possible in the present framework by so modifying the innovation processes as  to render the search greedier whilst retaining some features of the global search, especially the scrambling step. One way could be to start with \emph{pseudo-code 1}. In order to make the algorithm faster, we might replace the innovation used in \emph{pseudo-code 1} by the difference of the objective functional $f_{\text{exp}}$ computed over two succeeding iterations. Unfortunately, this leads to instability in the algorithm owing to the fact that the innovation becomes 1 dimensional in contrast to the number of design variables which is of order $O(10^4)$. This may be remedied by splitting the original objective functional into $101^2$ individual ones (i.e. one corresponding to each design variable) and thus constructing $101^2$ separate scalar innovation processes for the new objective function vector. This enables solving the problem in only a few iterations (as few as just 5 iterations) with only 30 MC particles. Accordingly, we now redefine  $f_{\text{exp}}^2=\sum\limits_{\varsigma=1}^{n}{{g_{\text{exp}}^{\varsigma}}}$ where $g_{\text{exp}}^{\varsigma}:=(\Phi_{\text{exp}}^\varsigma-\Phi^\varsigma)^2$.
   The present modification requires the realized innovation $\mathbf{I}_{i}$ at the $i^{\text{th}}$ iteration in \emph{pseudo-code 1} (specifically in step 3) to be given as:
  $$\mathbf{I}_{i}:=\left\{ \begin{matrix}
   {{{{g_{\text{exp}}^1}}}_{i-1}}\\...\\ {{{{g_{\text{exp}}^{n}}}}_{i-1}}\\\end{matrix}\right\}$$
It is recalled that this modified algorithm uses only scrambling for effecting the global search. This is an interesting example of the practical advantages in being able to suitably design the innovation process within the present framework. It is worth noting that our efforts at rendering the DE and the PSO greedier by tuning their appropriate algorithmic parameters have failed to make them effective in solving the current problem.

\begin{figure}[!h]
    \begin{center}
            \label{2}
           \includegraphics[width=0.8\textwidth]{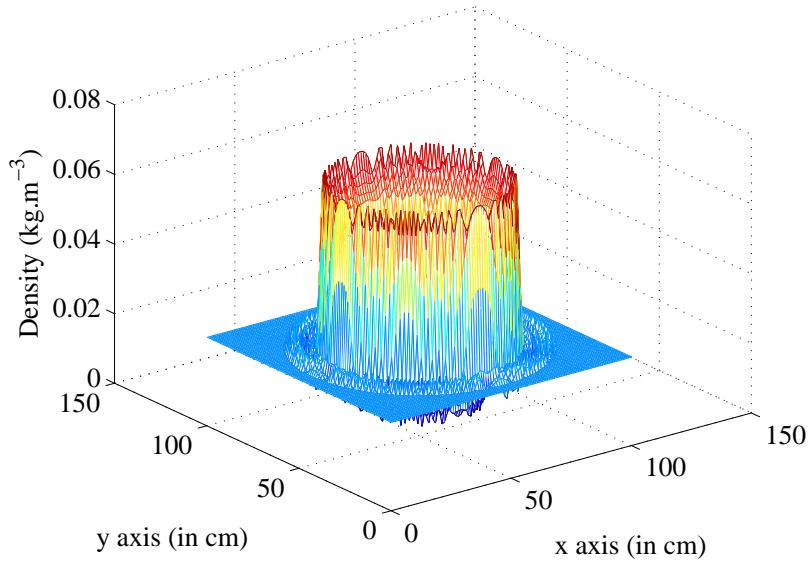}
    \end{center}
    \caption{%
        A 3D view of the reconstruction via the proposed framework
     }%
     \label{recon1}
\end{figure}

\begin{figure}[!h]
    \begin{center}
            \label{3}
           \includegraphics[width=0.8\textwidth]{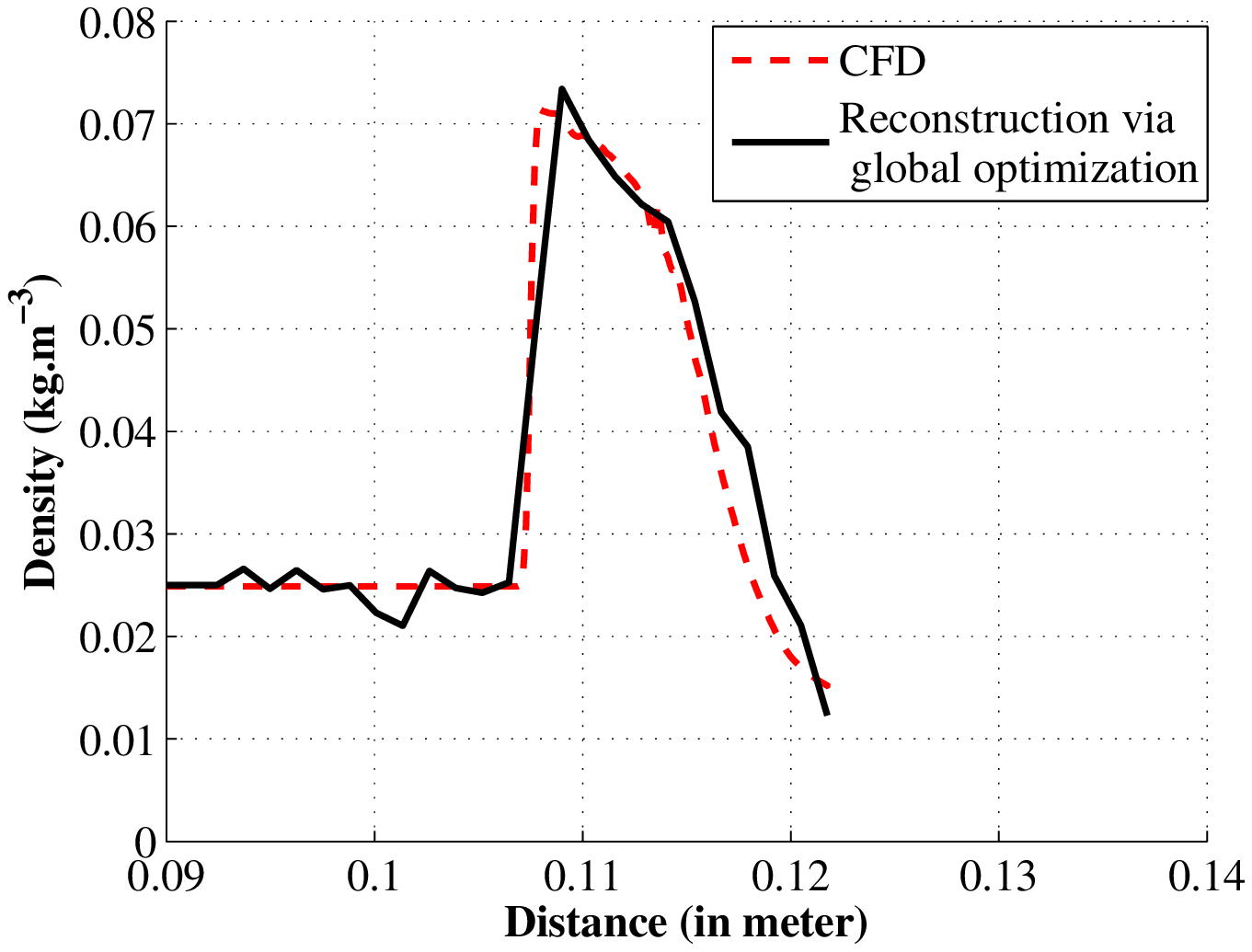}
    \end{center}
    \caption{%
        A projection of the reconstruction via the proposed framework
     }%
     \label{recon2}
\end{figure}

The recovered $\varrho${\bf(r)} is converted into density distribution using the Gladstone-Dale equation. A cross-section of the recovered density profile is shown in figures (\ref{recon1}) and (\ref{recon2}). The above reconstruction is verified (figure (\ref{recon2})) by computing the flow density distribution close to the boundary layer by solving Navier-Stoke's equation using the commercially available CFD software FLUENT. The comparison of the density variations near the boundary layer, shown in figure (\ref{recon2}), is seen to be quite close.
\section{Concluding remarks}
Despite the widely reported success and appealing simplicity of the methodological aspects of many existing evolutionary schemes for global optimization, often inspired by biological or social observations, the specific forms of their update scheme in approaching the global extremum are generally not rigorously derived. While a major aspect of this work is the attempt at finding an underlying logical thread that perhaps relates the basic ideas behind many such update strategies, the notion of a non-uniquely specifiable martingale approach in deriving the update term and its functional similarity with the directional derivative, used in most deterministic optimization schemes, has by far been the most fundamental aspect in this article. Such a derivative-free directional search has indeed been responsible for superior convergence features of the COMBEO family of schemes in contrast to some of the well known evolutionary methods. An effective global exploration of the search space however requires several random perturbation strategies, of which the coalescence step is treated as part of the martingale problem. It is shown that a variant of the coalescence step may mimic a PSO-like search. A DE-like search, on the other hand, may be organized through a variant of scrambling, which is yet another random perturbation strategy involving directional information swaps. The framework of COMBEO, founded on the theory of stochastic processes, yields highly competitive update schemes in order to meet the conflicting demands of faster convergence and efficacious exploration. As demonstrated via numerical work on a host of higher dimensional benchmark objective functions of the separable, partially separable and non-separable types, the COMBEO schemes show consistently superior performance in comparison to several existing evolutionary methods, prominently the DE and the PSO. The flexibility of the proposed optimization paradigm is also demonstrated by evolving a greedier version of the search scheme and applying the same to the reconstruction of boundary layer density variation in a high-speed flow experiment.
\par The generality of the framework afforded by COMBEO should motivate further research in improving the specific numerical schemes considered in this work. For instance, {\it{pseudo-codes 2}} and {\it{3}}, motivated respectively by the DE and the PSO, could possibly be rendered more competitive by incorporating some of the many available modifications or augmentations of the last two optimization methods. Work in this direction is currently under progress.

\appendix
\section{Derivation of the Extremal Equation}
\label{appendix}

Theorem 1:
For $\phi\in C_{b}^{2}$ the normalized conditional law ${{\pi }_{\tau }}\left({\phi}\right)$ of the solution process ${{\mathbf{x}}_{\tau}}$ satisfies the following extremal equation:

$$d{{\pi }_{\tau }}\left( \phi  \right)=\frac{1}{2}{{\pi }_{\tau }}\left( \sum\limits_{j,k=1}^{n}{\sum\limits
_{l=1}^{d}{\left( \frac{{{\partial }^{2}}\phi }{\partial {{x}^{j}}\partial {{x}^{k}}} \right){{g}^{jl}}{{g}
^{kl}}}} \right)d\tau+$$ $$\left( {{\pi }_{\tau }}\left( \phi h \right)-{{\pi }_{\tau }}\left( \phi  \right)
{{\pi }_{\tau }}\left( h \right) \right)\left( d{{{\overset{\lower0.5em\hbox{$\smash{\scriptscriptstyle\smile}
$}}{f}}}_{\tau}}-{{\pi}_{\tau}}\left(h\right)d\tau\right)$$


Proof:
The gain based update equation for a bounded and at least twice continuously differentiable
 function ${{{\mathbf{\phi}} }_{\tau }}:={\mathbf{\phi}} ({{\mathbf{x}}_{\tau }})$ of
 ${{\mathbf{x}}_{\tau }}$ may be arrived at by expanding $\phi ({{\mathbf{x}}_{\tau }})
{{\Lambda }_{\tau }}$, where $\tau \in \left( {{\tau }_{i-1}},{{\tau }_{i}} \right]$,
using Ito's formula:
\begin{equation}\label{A1}
d\left( {{\phi }_{\tau }}{{\Lambda }_{\tau }} \right)={{\phi }_{\tau }}d{{\Lambda }_{\tau }}
+d{{\phi }_{\tau }}{{\Lambda }_{\tau }}+\left\langle d{{\phi }_{\tau }},d{{\Lambda }_{\tau }}
 \right\rangle
\end{equation}
$\left\langle \cdot  \right\rangle $ denotes the quadratic covariation. A further expansion
leads to:

\begin{equation}\label{A2}
d\left( {{\phi }_{\tau }}{{\Lambda }_{\tau }} \right)={{\phi }_{\tau }}{{\Lambda }_{\tau }}
{{h}_{\tau }}d{{\overset{\lower0.5em\hbox{$\smash{\scriptscriptstyle\smile}$}}{f}}_{\tau}}+
{{\Lambda}_{\tau}}{{{\phi}'}_{\tau}}^{T}d{{\mathbf{x}}_{\tau}}+ \frac{1}{2}{{\Lambda }_{\tau }}
\left\langle d{{\mathbf{x}}_{\tau }},{{{{\phi }''}}_{\tau }}d{{\mathbf{x}}_{\tau }} \right\rangle
\end{equation}

By explicitly writing out the term $\left\langle d{{\mathbf{x}}_{\tau }},{{{{\phi }''}}
_{\tau }}d{{\mathbf{x}}_{\tau }} \right\rangle $ we get:

\begin{equation}\label{A3a}
\begin{split}
d\left( {{\phi }_{\tau }}{{\Lambda }_{\tau }} \right)={{\phi }_{\tau }}{{\Lambda }_{\tau }}
{{h}_{\tau }}d{{\overset{\lower0.5em\hbox{$\smash{\scriptscriptstyle\smile}$}}{f}}_{\tau}}
+\\{{\Lambda}_{\tau}}{{{\phi}'}_{\tau}}^{T}d{{\mathbf{x}}_{\tau}}+\frac{1}{2}\sum\limits_
{j,k=1}^{n}{\sum\limits_{l=1}^{d}{{{\left(\frac{{{\partial}^{2}}\phi}{\partial{{x}^{j}}
\partial{{x}^{k}}}\right)}_{\tau}}{{g}^{jl}}{{g}^{kl}}d\tau}}
\end{split}
\end{equation}

In deriving Eqn. (\ref{A3a}), Eqn. (\ref{process_eq}) is made use of. The incremental
form in Eqn. (\ref{A3a}) may be given the following integral representation:

\begin{equation} \label{A3b}
\begin{split}
{{\phi }_{\tau }}{{\Lambda }_{\tau }}={{\phi }_{i-1}}{{\Lambda }_{i-1}}+\int_{{{\tau }_
{i-1}}}^{\tau }{{{\Lambda }_{s}}{{\phi }_{s}}{{h}_{s}}d{{{\overset{\lower0.5em\hbox{$\smash{
\scriptscriptstyle\smile}$}}{f}}}_{s}}}+\\\int_{{{\tau}_{i-1}}}^{\tau}{{{\Lambda}_{s}}\left(
{{{{\phi}'}}_{s}}d{{\mathbf{\xi}}_{s}}+\frac{1}{2}\sum\limits_{j,k=1}^{n}{\sum\limits_{l=1}^
{d}{{{\left(\frac{{{\partial}^{2}}\phi}{\partial{{x}^{j}}\partial{{x}^{k}}}\right)}_{s}}{{g}
^{jl}}{{g}^{kl}}}}\right)ds}
\end{split}
\end{equation}

Taking conditional expectation with respect to ${{\mathcal{N}}_{\tau }}$ under $Q$, we get:

\begin{equation} \label{A4}
\begin{split}
{{E}_{Q}}\left( {{\phi }_{\tau }}{{\Lambda }_{\tau }}|{{\mathcal{N}}_{\tau }} \right)={{E}_{Q}}
\left( {{\phi }_{i-1}}{{\Lambda }_{i-1}}|{{\mathcal{N}}_{\tau }} \right)+\\{{E}_{Q}}\left(
\left( \int_{{{\tau }_{i-1}}}^{\tau }{{{\Lambda }_{s}}{{\phi }_{s}}{{h}_{s}}d{{{\overset{
\lower0.5em\hbox{$\smash{\scriptscriptstyle\smile}$}}{f}}}_{s}}}\right)|{{\mathcal{N}}_{\tau}}
\right)+\\{{E}_{Q}}\left(\left(\int\limits_{{{\tau}_{i-1}}}^{\tau}{{{\Lambda}_{s}}{{{{\phi}'}}
_{s}}d{{\mathbf{\xi}}_{s}}}\right)|{{\mathcal{N}}_{\tau}}\right)\\	 +\frac{1}{2}{{E}_{Q}}
\left( \left( \int_{{{\tau }_{i-1}}}^{\tau }{\sum\limits_{j,k=1}^{n}{\sum\limits_{l=1}^{d}{{{
\left( \frac{{{\partial }^{2}}\phi }{\partial {{x}^{j}}\partial {{x}^{k}}} \right)}_{s}}{{g}^
{jl}}{{g}^{kl}}}}ds} \right)|{{\mathcal{N}}_{\tau }} \right)		
\end{split}
\end{equation}
Using Fubini's theorem:

\begin{equation}\label{A5}
\begin{split}
{{E}_{Q}}\left( {{\phi }_{\tau }}{{\Lambda }_{\tau }}|{{\mathcal{N}}_{\tau }} \right)={{E}_{Q}}
\left( {{\phi }_{i-1}}{{\Lambda }_{i-1}}|{{\mathcal{N}}_{\tau }} \right)+
\int_{{{\tau }_{i-1}}}^{\tau }{{{E}_{Q}}\left( {{\Lambda }_{s}}{{\phi }_{s}}{{h}_{s}}|{{\mathcal{N}}
_{s}} \right)d{{{\overset{\lower0.5em\hbox{$\smash{\scriptscriptstyle\smile}$}}{f}}}_{s}}}+
\\\int\limits_{{{\tau}_{i-1}}}^{\tau}{{{E}_{Q}}\left({{\Lambda}_{s}}{{{{\phi}'}}_{s}}|{{\mathcal{N}}_{s}}
\right)d{{\mathbf{\xi}}_{s}}}
+\frac{1}{2}\int_{{{\tau }_{i-1}}}^{\tau }{{{E}_{Q}}\left( \sum\limits_{j,k=1}^{n}{\sum\limits_
{l=1}^{d}{{{\left( \frac{{{\partial }^{2}}\phi }{\partial {{x}^{j}}\partial {{x}^{k}}} \right)}_{
s}}{{g}^{jl}}{{g}^{kl}}}}|{{\mathcal{N}}_{s}} \right)}ds
\end{split}
\end{equation}
Noting that $\int\limits_{{{\tau }_{i-1}}}^{\tau }{{{E}_{Q}}\left( \left( {{\Lambda }_{s}}{{{{\phi
}'}}_{s}} \right)|{{\mathcal{N}}_{s}} \right)d{{\mathbf{\xi }}_{s}}}=0$ and for notational convenience
 denoting the un-normalized conditional expectation operator, ${{E}_{Q}}\left( {{\left( \centerdot
 \right)}_{\tau }}{{\Lambda }_{\tau }}|{{\mathcal{N}}_{s}} \right)$ as ${{\Theta }_{\tau }}\left(
\centerdot  \right)$ we arrive at the following equation:

\begin{equation}\label{A6}
\begin{split}
{{\Theta }_{\tau }}\left( \phi  \right)={{\Theta }_{i-1}}\left( \phi  \right)+\int_{{{\tau }_{i-1}}}
^{\tau }{{{\Theta }_{s}}\left( \phi h \right)d{{{\overset{\lower0.5em\hbox{$\smash{\scriptscriptstyle
\smile}$}}{f}}}_{s}}}+\frac{1}{2}\int_{{{\tau}_{i-1}}}^{\tau}{{{\Theta}_{s}}\left(\sum\limits_{j,k=1}
^{n}{\sum\limits_{l=1}^{d}{{{\left(\frac{{{\partial}^{2}}\phi}{\partial{{x}^{j}}\partial{{x}^{k}}}\right
)}_{s}}{{g}^{jl}}{{g}^{kl}}}}\right)ds}	
\end{split}
\end{equation}

An incremental representation of Eqn. (\ref{A6}) may be given as:

\begin{equation}\label{A7}
\begin{split}
d{{\Theta }_{\tau }}\left( \phi  \right)={{\Theta }_{\tau }}\left( \phi h \right)d{{\overset{\lower0.5em
\hbox{$\smash{\scriptscriptstyle\smile}$}}{f}}_{\tau}}+\frac{1}{2}{{\Theta}_{\tau}}\left(\sum\limits_
{j,k=1}^{n}{\sum\limits_{l=1}^{d}{\left(\frac{{{\partial}^{2}}\phi}{\partial{{x}^{j}}\partial{{x}^{k}}}
\right){{g}^{jl}}{{g}^{kl}}}}\right)d\tau	
\end{split}
\end{equation}

In order to obtain the normalized conditional law, i.e. ${{\pi }_{\tau }}\left( \phi  \right)=\frac{{{\Theta }_{\tau }}\left( \phi  \right)}{{{\Theta }_{\tau }}\left( 1 \right)}$, it is expanded using Ito's
formula as given below:

\begin{equation}\label{A8}
\begin{split}
d{{\pi }_{\tau }}\left( \phi  \right)=\frac{d{{\Theta }_{\tau }}\left( \phi  \right)}{{{\Theta }_{\tau }}
\left( 1 \right)}+{{\Theta }_{\tau }}\left( \phi  \right)d\left( \frac{1}{{{\Theta }_{\tau }}\left( 1
\right)} \right)+\left\langle d{{\Theta }_{\tau }}\left( \phi  \right),d\left( \frac{1}{{{\Theta }_{
\tau }}\left( 1 \right)} \right) \right\rangle	
\end{split}
\end{equation}

$d\left( \frac{1}{{{\Theta }_{\tau }}\left( 1 \right)} \right)$ may be expanded as:

\begin{equation}\label{A9}
d\left( \frac{1}{{{\Theta }_{\tau }}\left( 1 \right)} \right)=-\frac{1}{\Theta _{\tau }^{2}\left( 1 \right)
}d{{\Theta }_{\tau }}\left( 1 \right)+\frac{1}{\Theta _{\tau }^{3}\left( 1 \right)}\left\langle d{{\Theta }_
{\tau }}\left( 1 \right),d{{\Theta }_{\tau }}\left( 1 \right) \right\rangle
\end{equation}

Putting $\phi =1$ in Eqn. (\ref{A7}), we get an Ito expansion for $d{{\Theta }_{\tau }}\left( 1 \right)$,
which is given below:

\begin{equation}\label{A10}
d{{\Theta }_{\tau }}\left( 1 \right)={{\Theta }_{\tau }}\left( h \right)d{{\overset{\lower0.5em\hbox{$\smash{\scriptscriptstyle\smile}$}}{f}}_{\tau}}	 
\end{equation}

\par Using Eqn. (\ref{A10}) in Eqn. (\ref{A9}):

\begin{equation}\label{A11}
d\left( \frac{1}{{{\Theta }_{\tau }}\left( 1 \right)} \right)=-\frac{{{\pi }_{\tau }}\left( h \right)}{{{
\Theta }_{\tau }}\left( 1 \right)}d{{\tilde{f}}_{\tau }}+\frac{\pi _{\tau }^{2}\left( h \right)}{{{\Theta
}_{\tau }}\left( 1 \right)}d\tau
\end{equation}	
\par Using Eqn. (\ref{A7}) and (\ref{A11}) in Eqn. (\ref{A8}) we get:

\begin{equation}\label{A12}
\begin{split}
d{{\pi }_{\tau }}\left( \phi  \right)={{\pi }_{\tau }}\left( \phi h \right)d{{\overset{\lower0.5em\hbox
{$\smash{\scriptscriptstyle\smile}$}}{f}}_{\tau}}+\frac{1}{2}{{\pi}_{\tau}}\left(\sum\limits_{j,k=1}^{n}
{\sum\limits_{l=1}^{d}{\left(\frac{{{\partial}^{2}}\phi}{\partial{{x}^{j}}\partial{{x}^{k}}}\right){{g}^{jl}}
{{g}^{kl}}}}\right)d\tau+\\\left( -{{\pi }_{\tau }}\left( \phi  \right){{\pi }_{\tau }}\left( h \right)d{{{
\overset{\lower0.5em\hbox{$\smash{\scriptscriptstyle\smile}$}}{f}}}_{\tau}}+{{\pi}_{\tau}}\left(\phi\right)
\pi_{\tau}^{2}\left(h\right)d\tau\right)-{{\pi }_{\tau }}\left( \phi h \right){{\pi }_{\tau }}\left( h \right)d\tau
\end{split}
\end{equation}

Thus we arrive at the extremal equation for the evolution of the normalized conditional estimate ${{\pi }_
{\tau }}\left( \phi  \right)$ given as:

\begin{equation}\label{A13}
\begin{split}
d{{\pi }_{\tau }}\left( \phi  \right)=\frac{1}{2}{{\pi }_{\tau }}\left( \sum\limits_{j,k=1}^{n}{\sum\limits
_{l=1}^{d}{\left( \frac{{{\partial }^{2}}\phi }{\partial {{x}^{j}}\partial {{x}^{k}}} \right){{g}^{jl}}{{g}
^{kl}}}} \right)d\tau\\+\left( {{\pi }_{\tau }}\left( \phi h \right)-{{\pi }_{\tau }}\left( \phi  \right)
{{\pi }_{\tau }}\left( h \right) \right)\left( d{{{\overset{\lower0.5em\hbox{$\smash{\scriptscriptstyle\smile}
$}}{f}}}_{\tau}}-{{\pi}_{\tau}}\left(h\right)d\tau\right)
\end{split}
\end{equation}

Remark 1: Eqn. (\ref{A6}) finds its equivalence with the Zakai equation well known in stochastic filtering.
\par Remark 2: Eqn. (\ref{A13}) has its parallel in the Kushner-Stratonovich equation, which is again well known in stochastic
filtering.

Theorem 2:
 When $\phi\left({{x}}\right)=x$ Eqn. {\ref{A13}} may be rewritten as:

$$d{{\pi }_{\tau }}\left({\mathbf{x}} \right)=\left( {{\pi }_{\tau }}\left({\mathbf{x}}f \right)-{{\pi }_{\tau }}
\left({\mathbf{x}} \right){{\pi }_{\tau }}\left( f \right) \right){{\left( {{\rho }_{\tau }}\rho _{\tau }^{T}
 \right)}^{-1}}\left( d{{{\overset{\lower0.5em\hbox{$\smash{\scriptscriptstyle\smile}$}}{f}}}_{\tau}}-{{\pi}
_{\tau}}\left(f\right)d\tau\right)$$

Proof:
 Since we are typically interested in the evolution of the conditional estimate of ${{\mathbf{x}}_
{\tau }}$, i.e. $\phi $ is the identity function, Eqn. (\ref{A13}) may be simplified as:
\begin{equation}\label{A14}
d{{\pi }_{\tau }}\left( {\mathbf{x}} \right)=\left( {{\pi }_{\tau }}\left( {\mathbf{x}}h \right)-{{\pi }_{\tau
 }}\left( {\mathbf{x}} \right){{\pi }_{\tau }}\left( h \right) \right)\left( d{{{\overset{\lower0.5em\hbox{$\smash{\scriptscriptstyle\smile}$}}{f}}}_{\tau}}-{{\pi}_{\tau}}\left(h\right)d\tau\right)	
\end{equation}

Replacing $h({{\mathbf{x}}_{\tau }})$ by $\rho _{\tau }^{-1}f({{\mathbf{x}}_{\tau }})$ in Eqn. (\ref{A14}), we get:

\begin{equation}\label{A15}
\begin{split}
d{{\pi }_{\tau }}\left({\mathbf{x}} \right)=\left( {{\pi }_{\tau }}\left({\mathbf{x}}f \right)-{{\pi }_{\tau }}
\left({\mathbf{x}} \right){{\pi }_{\tau }}\left( f \right) \right){{\left( {{\rho }_{\tau }}\rho _{\tau }^{T}
 \right)}^{-1}}\left( d{{{\overset{\lower0.5em\hbox{$\smash{\scriptscriptstyle\smile}$}}{f}}}_{\tau}}-{{\pi}
_{\tau}}\left(f\right)d\tau\right)
\end{split}
\end{equation}

\end{document}